\documentclass[a4paper, 12pt, numbers = noenddot, parskip = no]{scrartcl}
\usepackage[T1]{fontenc}
\usepackage[utf8]{inputenc}
\usepackage{lmodern}
\usepackage{lscape}
\usepackage{siunitx}
\setkomafont{sectioning}{\normalfont \rmfamily \bfseries}%
\setkomafont{paragraph}{\normalfont \rmfamily \bfseries}%

\usepackage{url}
\usepackage{hyperref}%
\hypersetup{%
  breaklinks=true,%
  linktocpage=true,
  colorlinks=true,%
  linkcolor=black,%
  citecolor=blue,%
  filecolor=blue,%
  urlcolor=blue,%
  pdftitle={},%
  pdfsubject={},%
  pdfauthor={},%
  pdfkeywords={},%
  pdfproducer={}%
  }%

\usepackage{mathtools} 
\mathtoolsset{showonlyrefs=true}
\usepackage[ntheorem, framemethod=TikZ]{mdframed} 
\usepackage{framed} 
\usepackage[amsmath, hyperref, framed]{ntheorem}%
\usepackage{amssymb} 
\usepackage{bm}
\usepackage{wasysym}

\usepackage{algorithm}
\usepackage[noend]{algpseudocode}

\newcommand{\testFun}{f}%

\newcommand{\ess}{\mathop{\mathrm{ess}}}

\newcommand{\sta}{x}
\newcommand{\Sta}{X}
\newcommand{\staAll}{\mathbf{\sta}}
\newcommand{\StaAll}{\mathbf{\Sta}}

\newcommand{\StaRes}{\widetilde{X}}

\newcommand{\StaResAll}{\widetilde{\mathbf{\Sta}}}
\newcommand{\anc}{a}
\newcommand{\Anc}{A}
\newcommand{\ancAll}{\mathbf{\anc}}
\newcommand{\AncAll}{\mathbf{\Anc}}
\newcommand{\aux}{u}
\newcommand{\Aux}{U}

\newcommand{\wei}{w}
\newcommand{\Wei}{W}
\newcommand{\WeiAll}{\mathbf{\Wei}}

\newcommand{\weiRes}{\widetilde{w}}
\newcommand{\WeiRes}{\widetilde{W}}
\newcommand{\weiResAll}{\widetilde{\mathbf{\wei}}}

\newcommand{\resDist}{\rho}
\newcommand{\resDistAlt}{\varrho}
\newcommand{\indDist}{\lambda}
\newcommand{\indDistAll}{\indDist}
\newcommand{\indDistAlt}{\kappa}
\newcommand{\indDistAllAlt}{\indDistAlt}
\newcommand{\weiResFun}{\phi}

\newcommand{\resScheme}{\langle \resDist, \indDistAll\rangle}
\newcommand{\resSchemeNum}[1]{\langle \resDist, \indDistAll_{#1}\rangle}
\newcommand{\resSchemeTime}[1]{\langle \resDist_{#1-1}, \indDistAll_{#1}\rangle}
\newcommand{\resSchemeAlt}{\langle \resDistAlt, \indDistAllAlt\rangle}

\newcommand{\target}{\pi}
\newcommand{\targetEst}{\hat{\target}}
\newcommand{\targetEstRes}{\tilde{\target}}

\newcommand{\smc}{\mathrm{smc}}
\newcommand{\csmc}{\mathrm{csmc}}
\newcommand{\csmcWithAncestorSampling}{\mathrm{csmc}^*}
\newcommand{\ancestorTracing}{\mathrm{anc}}
\newcommand{\backwardSampling}{\mathrm{back}}
\newcommand{\backwardKernel}{B}

\newcommand{\permutations}{\mathsf{S}}
\newcommand{\cycles}{\mathsf{C}}

\newcommand{\normConst}{\mathcal{Z}} 

\DeclareMathOperator{\dN}{N} 
\DeclareMathOperator{\dCat}{Cat} 
\DeclareMathOperator{\dUnif}{Unif} 

\newcommand{\diff}{\mathrm{d}} 
\DeclareMathOperator{\Prob}{\mathbb{P}} 
\DeclareMathOperator{\E}{\mathbb{E}} 
\DeclareMathOperator{\var}{var} 
\DeclareMathOperator{\ind}{\mathbf{1}} 
\newcommand{\ccdot}{\,\cdot\,} 
\DeclareMathOperator{\bo}{\mathcal{O}}%
\DeclareMathOperator{\vol}{vol}%
\newcommand{\iidSim}{\mathrel{\overset{\mathsc{iid}}{\sim}}}

\newcommand{\calA}{\mathcal{A}}%
\newcommand{\calB}{\mathcal{B}}%
\newcommand{\calF}{\mathcal{F}}%
\newcommand{\calX}{\mathcal{X}}%
\newcommand{\mathsc}[1]{{\normalfont\textsc{#1}}}%

\newcommand{\reals}{\mathbb{R}}%
\newcommand{\naturals}{\mathbb{N}}%
\newcommand{\spaceU}{\mathsf{U}}%
\newcommand{\spaceX}{\mathsf{X}}%
\newcommand*{\mycmss}{\fontfamily{cmss}\selectfont}
\DeclareTextFontCommand{\textcmss}{\mycmss}

\newcommand{\prodSubstackAligned}[6]{
  \prod_{\substack{\mathllap{#1} #2 \mathrlap{#3}\\
                   \mathllap{#4} #5 \mathrlap{#6}}}}%
                   
\usepackage{graphicx,accents}

\makeatletter
\DeclareRobustCommand{\cev}[1]{%
  \mathpalette\do@cev{#1}%
}
\newcommand{\do@cev}[2]{%
  \fix@cev{#1}{+}%
  \reflectbox{$\m@th#1\vec{\reflectbox{$\fix@cev{#1}{-}\m@th#1#2\fix@cev{#1}{+}$}}$}%
  \fix@cev{#1}{-}%
}
\newcommand{\fix@cev}[2]{%
  \ifx#1\displaystyle
    \mkern#23mu
  \else
    \ifx#1\textstyle
      \mkern#23mu
    \else
      \ifx#1\scriptstyle
        \mkern#22mu
      \else
        \mkern#22mu
      \fi
    \fi
  \fi
}
\makeatother

\makeatletter
\def\theorem@checkbold{}
\makeatother

\newcommand{\qedwhite}{\hfill \ensuremath{\Box}}

\usepackage{scalerel,stackengine}
\makeatletter
\stackMath
\newcommand\reallywidehat[1]{%
\savestack{\tmpbox}{\stretchto{%
  \scaleto{%
    \scalerel*[\widthof{\ensuremath{#1}}]{\kern.1pt\mathchar"0362\kern.1pt}%
    {\rule{0ex}{\textheight}}
  }{\textheight}%
}{2.4ex}}%
\stackon[-6.9pt]{#1}{\tmpbox}%
}
\makeatother

\usepackage{graphicx}
\DeclareGraphicsExtensions{.png, .pdf} 
\usepackage{float}
\usepackage{caption, subcaption}

\usepackage{xcolor} 

\usepackage{tikz}

\qedsymbol{\ensuremath{\Box}}
\theoremstyle{plain}%

\theoremheaderfont{\normalfont\bfseries\rmfamily} 
\theorembodyfont{\normalfont\slshape}%
\theoremseparator{.}
\newtheorem{proposition}{Proposition}%
\newtheorem{lemma}{Lemma}%
\newtheorem{assumption}{Assumption}%
\newtheorem{definition}{Definition}%
\newtheorem{remark}{Remark}%
\newtheorem{example}{Example}%

\theorembodyfont{\rmfamily}%
\newmdtheoremenv[
 ntheorem=true,
 skipbelow = .5\baselineskip plus 1ex minus 1ex,
 skipabove = .5\baselineskip plus 1ex minus 1ex,
 innerleftmargin = 0pt,
 innerrightmargin = 0pt,
 leftline = false,
 rightline = false,
 needspace = 0ex 
]{framedAlgorithm}[theorem]{Algorithm}

\newmdtheoremenv[
 ntheorem=true,
 skipbelow = .5\baselineskip plus 1ex minus 1ex,
 skipabove = .5\baselineskip plus 1ex minus 1ex,
 innerleftmargin = .4\baselineskip plus 1ex minus 1ex,
 innerrightmargin = .4\baselineskip plus 1ex minus 1ex,
 leftline = true,
 rightline = true,
 needspace = 0ex 
]{framedExample}[theorem]{Application to state-space models}

\theoremstyle{empty}%
\theoremheaderfont{} 

\theoremstyle{nonumberplain}%
\theorembodyfont{\upshape \rmfamily}%
\theoremheaderfont{\upshape \rmfamily\bfseries}%
\theoremsymbol{\ensuremath{_\Box}}
\newtheorem{namedProof}{Proof}%

\usepackage[textwidth=2.5cm, textsize=scriptsize]{todonotes}

\usepackage{xcolor}
\definecolor{colEdit}{HTML}{00BFC4}

\usepackage[inline]{enumitem}

\newlist{myenumerate}{enumerate}{3} 
\setlist[myenumerate]{itemsep=1ex, topsep=0.7ex}
\setlist[myenumerate,1]{label=\arabic*., ref=\arabic*, leftmargin=*, itemsep=-0.1ex, topsep=0.7ex}
\setlist[myenumerate,2]{label=\roman*., ref=\roman*, labelindent=\parindent, topsep=-0.25ex, itemsep=-0.1ex}
\setlist[myenumerate,3]{label=\alph*., ref=\alph*, labelindent=\parindent, topsep=-0.25ex, itemsep=-0.1ex}
\newlist{myitemize}{enumerate}{2} 
\setlist[myitemize]{itemsep=1ex, topsep=0.5ex}
\setlist[myitemize,1]{label=\textbullet, leftmargin=*, itemsep=-0.1ex, topsep=0.6ex}
\setlist[myitemize,2]{label=$-$, labelindent=\parindent, topsep=-0.25ex, itemsep=-0.1ex}

\makeatletter
\newcommand{\myitem}[1]{%
\item[#1.]\protected@edef\@currentlabel{#1}%
}
\makeatother

\usepackage{ifthen}%
\usepackage{mfirstuc}
\usepackage{xkeyval}%
\usepackage{xfor}%
\usepackage{amsgen}%
\usepackage{etoolbox}%
\usepackage{datatool-base}%

\usepackage[%
  xindy,%
  style=long,%
  toc=true,%
  nonumberlist,%
  acronym,%
  acronymlists={abbreviations},%
  hyperfirst=true
  ]{glossaries}%
  
\newacronym{PMCMC}{PMCMC}{particle Markov chain Monte Carlo}%
\newacronym{SIR}{SIR}{sampling--importance resampling}%
\newacronym{EMCMC}{EMCMC}{ensemble Markov chain Monte Carlo}%
\newacronym{EHMM}{EHMM}{embedded hidden Markov model}%
\newacronym{HMM}{HMM}{hidden Markov model}%
\newacronym{CPFBS}{CPF-BS}{conditional particle filter with backward sampling}%
\newacronym{CPFAS}{CPF-AS}{conditional particle filter with backward sampling}%
\newacronym{PG}{PG}{particle Gibbs}%
\newacronym{PGBS}{PG-BS}{particle Gibbs sampler with backward sampling}%
\newacronym{PGAS}{PG-AS}{particle Gibbs sampler with backward sampling}%
\newacronym{SQMC}{SQMC}{sequential quasi Monte Carlo}%
\newacronym{RQMC}{RQMC}{randomised quasi Monte Carlo}%
\newacronym[user1={ancestor-sampling}]{AS}{AS}{ancestor sampling}%
\newacronym[user1={backward-sampling}]{BS}{BS}{backward sampling}%
\newacronym{PDF}{PDF}{probability density function}%
\newacronym{IID}{IID}{independent and identically distributed}%
\newacronym{MCMC}{MCMC}{Markov chain Monte Carlo}%
\newacronym{MH}{MH}{Metropolis--Hastings}%
\newacronym{ESS}{ESS}{effective sample size}%
\newacronym{SNIS}{SNIS}{self-normalised importance sampling}%
\newacronym{PMMH}{PMMH}{particle marginal Metro\-po\-lis--Has\-tings}%
\newacronym{MCWM}{MCWM}{Monte Carlo within Metropolis}%
\newacronym{CDF}{CDF}{cumulative distribution function}%
\newacronym{SMC}{SMC}{sequential Monte Carlo}%
\newacronym{CSMC}{CSMC}{conditional sequential Monte Carlo}%
\newacronym{EPSRC}{EPSRC}{Engineering and Physical Sciences Research Council}%
\newacronym{ESJD}{ESJD}{expected squared jumping distance}%

\newacronym{PF}{PF}{particle filter}%
\newacronym{BPF}{BPF}{bootstrap PF}%
\newacronym{FAAPF}{FAAPF}{fully-adapted auxiliary PF}%
\newacronym{APF}{APF}{general auxiliary PF}%

\newacronym{MPF}{m-PF}{marginal particle filter}%
\newacronym{BMPF}{m-BPF}{marginal BPF}%
\newacronym{FAAMPF}{m-FAAPF}{marginal FAAPF}%
\newacronym{AMPF}{m-APF}{marginal APF}%

\newacronym{MCMCPF}{MCMC-PF}{MCMC-PF}%
\newacronym{MCMCBPF}{MCMC-BPF}{MCMC-BPF}%
\newacronym{MCMCFAAPF}{MCMC-FA-APF}{MCMC-FAAPF}%
\newacronym{MCMCAPF}{MCMC-APF}{MCMC-APF}%

\newacronym{MCMCMPF}{MCMC-m-PF}{MCMC m-PF}%
\newacronym{MCMCBMPF}{MCMC-m-BPF}{MCMC-m-BPF}%
\newacronym{MCMCFAAMPF}{MCMC-m-FAAPF}{MCMC-m-FAAPF}%
\newacronym{MCMCAMPF}{MCMC-m-APF}{MCMC-m-APF}%

\newacronym{CPF}{CPF}{conditional PF}%
\newacronym{ICPF}{ICPF}{iterated CPF}%
\newacronym{ICPFBS}{ICPF-BS}{ICPF with backward sampling}%
\newacronym{ICPFAS}{ICPF-AS}{ICPF with ancestor sampling}%
\newacronym{CMPF}{m-CPF}{marginal conditional PF}%
\newacronym{ICMPF}{m-ICPF}{marginal iterated CPF}%

\usepackage{csquotes}%
\usepackage{natbib}
\setcitestyle{authoryear}
\newcommand{\algorithmSepAbove}{\vspace{-2.5ex}}
\newcommand{\algorithmSepBelow}{\vspace{-2.5ex}}
\newcommand{\algorithmSepAboveAlt}{\vspace{-2.5ex}}
\newcommand{\algorithmSepAboveSmall}{\vspace{-1ex}}
\newcommand{\algorithmSepBelowSmall}{\vspace{-1ex}}

\title{Resampling in conditional SMC algorithms}
\author{
  Axel Finke\thanks{School of Mathematics, Statistics and Physics, Newcastle University, Newcastle upon Tyne, UK. \texttt{axel.finke@ncl.ac.uk}. ORCID: 0000-0002-8379-3012} \and
  Adam M.\ Johansen\thanks{Department of Statistics, University of Warwick, Coventry, UK. \texttt{a.m.johansen@warwick.ac.uk}. ORCID: 0000-0002-3531-7628} \and
  Anthony Lee\thanks{School of Mathematics, University of Bristol, Bristol, UK. \texttt{anthony.lee@bristol.ac.uk}. ORCID: 0000-0001-7765-0616} \and
  Lawrence M.\ Murray\thanks{Independent researcher. \texttt{lawrence@indii.org}. ORCID: 0000-0002-6567-6015}
}
\date{\today}

\begin{document}

\maketitle

\begin{abstract}
\noindent Conditional sequential Monte Carlo (CSMC) algorithms arise in particle Markov chain Monte Carlo and a number of related settings. As in standard sequential Monte Carlo (SMC) algorithms, it is possible to employ a number of approaches to resampling within CSMC, but some additional care is required to arrive at a valid algorithm. We present a simple framework for implementing valid SMC and CSMC algorithms which (a) covers most known resampling schemes including those with a complicated dependence structure like systematic resampling, but also adaptive resampling, and even more `exotic' schemes like a version of chopthin resampling; (b) explains how to implement conditional analogues of these and other well known resampling schemes without randomly permuting/shifting the order of the ancestor indices; (c) requires only very weak assumptions which include neither (marginal) `unbiasedness' nor exchangeability.

\end{abstract}

\section{Introduction}

\subsection{Resampling}

\emph{\Gls{SMC}} algorithms (also termed \emph{particle filters}) are a class of Monte Carlo methods for approximating (sequences of) probability measures and their normalising constants such as marginal likelihoods \citep[see, e.g.,][and references therein]{chopin2020introduction}. The closely related \emph{\gls{CSMC}} algorithm induces a Markov chain which can be used to efficiently approximate the joint posterior distribution of all the latent states in state-space models (a.k.a., the \emph{joint smoothing distribution}), and more generally, due to its ability to exploit the `decorrelation-over-time' structure of the target distribution \citep{andrieu2010particle, lee2020coupled, karjalainen2025mixing}. \gls{CSMC} methods are also at the core of more complex schemes, e.g., for unbiased estimation \citep{jacob2019smoothing, boom2022unbiased, shestopaloff2019replica} or diffusion models \citep{corenflos2024conditioning, stanton2024particle}. 

A key ingredient in \gls{SMC} algorithms is resampling and \gls{CSMC} algorithms rely on a related \emph{conditional} resampling mechanism. Numerous resampling schemes have been proposed, e.g., multinomial resampling \citep{rubin1987calculation}, stratified resampling \citep{kitagawa1996monte}, systematic resampling \citep{carpenter1999improved}, residual resampling \citep{liu1998sequential, douc2005comparison}, adaptive resampling \citet{liu1995blind}, killing resampling \citep{delmoral2013mean, chopin2022resampling}, Metropolis resampling \citep{murray2016parallel}, Hilbert-curve resampling \citep{gerber2015sequential}, chopthin resampling \citep{gandy2015chopthinv3}, 
as well as further sophisticated resampling schemes in \citet{gerber2019negative, chopin2022resampling}. For practitioners, it is therefore important to have verifiable conditions which guarantee that their chosen resampling scheme leads to \gls{SMC} or \gls{CSMC} algorithms that are valid. In this work, we introduce a perspective which allows us to view most of these algorithms within a common framework which provides such guarantees.

\subsection{Existing frameworks}

We will consider `validity' of \gls{SMC} and \gls{CSMC} algorithms only in the sense that the former yield unbiased estimates of normalising constants (e.g., marginal likelihoods) and that the latter induce Markov kernels which leave the target distribution (e.g., joint smoothing distribution) invariant. We do not analyse the efficiency (e.g.\ convergence rates) associated with different resampling schemes. For the latter, we refer the reader to \citet{douc2005comparison, whiteley2016role, gerber2019negative, webber2019unifying, chopin2022resampling, li2022stratification}.

When homogeneous (i.e., equal) post-resampling weights are sought, a common requirement used to ensure validity (in the above-mentioned sense) is the conventional \emph{`unbiasedness'} condition from \citet{crisan1998discrete} which says that the expected number of descendants of each particle should be proportional to its weight.
  
For \gls{CSMC} algorithms, the seminal work of \citet{andrieu2010particle} additionally required that resampling schemes should be exchangeable. They also noted that exchangeability can always be ensured by randomly permuting the order of the ancestor indices after resampling. This idea was used in \citet{chopin2015particle} to derive a conditional version of residual-multinomial resampling. However, such random permutations can incur a non-negligible computational cost, especially on parallel architectures \citep{murray2016parallel}, and the framework from \citet{andrieu2010particle} does not cover adaptive resampling.
  
Alternatively, \citet{chopin2015particle, karppinen2023conditional} imposed a so-called \emph{marginal `unbiasedness'} condition which implies the conventional `unbiasedness' condition but does not require exchangeability. For some non-exchangeable resampling schemes (i.e., systematic resampling and killing resampling) the authors were thus able to derive a valid \gls{CSMC} algorithm by applying a random shift to the order of the ancestor indices after resampling. However, the random shift again incurs additional computational cost and their framework does not cover adaptive resampling.

It has long been known that the conventional `unbiasedness' condition is not necessary and can be dropped at the expense of non-uniform post-resampling weights \citep[e.g.,][]{liu2001monte}. \citet{chen2010particle} justified this strategy as changing the sequence of intermediate distributions of the \gls{SMC} or \gls{CSMC} algorithm. However, this justification does not apply to non-exchangeable resampling schemes such as the partial resampling scheme from \citet{martino2016weighting} or to the optimal finite-state resampling scheme from \citet{fearnhead2003online}.

Finally, \citet{whiteley2016role, huggins2019sequential} proved the validity of \gls{SMC} and \gls{CSMC} algorithms for (adaptive) resampling schemes that fit into their $\alpha$\gls{SMC} framework. However, this framework's resampling-matrix representation only exists for resampling schemes in which the ancestor indices are conditionally independent and thus does not apply to the widely used systematic resampling scheme nor to Metropolis resampling or Hilbert-curve resampling.

\subsection{Contributions}

We show how to construct valid \gls{SMC} and \gls{CSMC} algorithms based on a wide range of resampling schemes and in particular:
\begin{enumerate}

  \item In Section~\ref{sec:framework}, we present a simple framework for resampling which covers most known resampling schemes and their conditional versions as special cases, including those with a complicated dependence structure like (conditional) systematic resampling and those which do not satisfy the conventional `unbiasedness' condition from \citet{crisan1998discrete} such as partial resampling \citep{martino2016weighting}, optimal finite-state resampling scheme \citep{fearnhead2003online} or a version of chopthin resampling \citep{gandy2015chopthinv3}.

  \item  Key to our framework is the introduction of an \emph{index distribution} $\indDist(\ccdot|n)$ which governs the law of the particle index of the descendant of particle~$n$ on a suitably extended space. By making the r\^ole and the (potentially non-uniform) choice of this index distribution explicit, our work automatically provides conditional versions of any resampling scheme within our framework. For example, our work yields a novel conditional version of chopthin resampling. Further, our work makes it possible to implement conditional versions of non-exchangeable resampling schemes without having to randomly permute or shift the order of the ancestor indices.
  
  \item The assumptions of our framework are weaker than those imposed in the existing frameworks from \citet{andrieu2010particle, chopin2015particle, karppinen2023conditional, whiteley2016role, huggins2019sequential}, requiring only a mild \emph{persistence} condition (Definition~\ref{def:persistence}) which ensures that the post-resampling weights cannot all be simultaneously zero and a \emph{proper weighting} assumption (Definition~\ref{def:proper_weighting}). Additionally, we specify \emph{balanced} index distribution (Definition~\ref{def:balance}) which automatically ensures proper weighting. Section~\ref{sec:examples} shows that most known resampling schemes implicitly use a balanced index distribution.
  
  \item In Section~\ref{sec:application_to_smc_methods}, we prove that our framework makes it simple to guarantee (a) the unbiasedness of normalising-constant estimates generated by \gls{SMC} algorithms; (b) the validity of the corresponding \gls{CSMC} algorithm (with backward/ancestor sampling), for a wide range of resampling schemes---including adaptive/systematic resampling. In Appendix~\ref{app:sec:on_naive_implementations_of_conditional_resampling}, we further demonstrate that na\"ive implementations of conditional resampling sometimes used by practitioners can bias the Markov chain induced by the \gls{CSMC} algorithm and may even worsen its autocorrelation.
\end{enumerate}

\subsection{Notation}

We let $[n] \coloneqq \{1,\dotsc,n\}$, for any $n \in \naturals \coloneqq \{1, 2, \dotsc\}$ and use the convention $\tfrac{0}{0} \coloneqq 0$. Throughout this work, we assume some background probability space $(\varOmega, \calA, \Prob)$; $\mathcal{B}(\spaceX)$ denotes the Borel $\sigma$-algebra on some topological space $\spaceX$. For any measure space $(\spaceX, \calX, \mu)$, where $\mu$ may be random, we write $\smash{\mu(f) \coloneqq \int_\spaceX f(x) \mu(\diff x)}$, for $\mu$-integrable functions $f \colon \spaceX \to \reals$. Abusing notation, we frequently use the same symbol for measures and their densities (w.r.t.\ a suitable dominating measure, e.g., Lebesgue or counting measure); $\vol(A)$ is the volume (i.e., Lebesgue measure) of measurable $A \in \reals^d$; $\smash{\dCat(p_{1:K})}$ denotes the categorical distribution on $[K]$ with probability vector $\smash{p_{1:K} \in [0,1]^K}$ such that $\smash{\sum_{k=1}^K p_k = 1}$. For a suitable set $B$, $\dUnif_B$ is the (discrete or continuous) uniform distribution on $B$; $\delta_x(B) = \ind\{x \in B\}$ is the point mass/Dirac measure at $x$; if $B = \{b\}$ is a singleton, we write $\smash{\delta_x(\{b\}) =: \delta_x(b)}$.

\section{Framework for the validity of resampling schemes}
\label{sec:framework}

\subsection{Resampling}

\subsubsection{Goal of resampling}

In this section, we present a generic framework for (conditional) resampling. We acknowledge that this necessitates a level of abstraction whose utility will become fully apparent in Section~\ref{sec:examples}, where we illustrate how most known resampling schemes fit into our framework. Due to the notational burden of \gls{SMC} and \gls{CSMC} algorithms, we defer these methods to Section~\ref{sec:application_to_smc_methods} and treat (conditional) resampling in a simpler setting until then.

Hereafter, assume some measurable space $(\spaceX, \calB(\spaceX))$ where $\spaceX$ is Polish, e.g., $\smash{\spaceX = \reals^d}$. Consider a (possibly random)  weighted sample $(\StaAll, \WeiAll) \coloneqq (\Sta^{1:N}, \Wei^{1:N})$, for some $N \in \naturals$. Here, for all $n \in [N]$, $\Sta^n$ is a sample point (`particle') that takes values in $\spaceX$ and $\Wei^n$ is a non-negative weight, such that $\sum_{n=1}^N \Wei^n = 1$. Such a weighted sample can be associated with a probability measure on $(\spaceX, \calB(\spaceX))$:
\begin{align}
  \targetEst \coloneqq \sum_{n=1}^N \Wei^n \delta_{\sta^n}.
\end{align}
Resampling is a procedure that replaces $(\StaAll, \WeiAll)$ with random variables $(\StaResAll, \weiResAll) \coloneqq (\StaRes^{1:M}, \weiRes^{1:M})$, for some $M \in \naturals$, which can be associated with the random measure:
\begin{align}
  \targetEstRes \coloneqq \sum_{m=1}^M \weiRes^m \delta_{\StaRes^m}.
\end{align}
A typical goal of resampling is to ensure that $\targetEstRes$ is close to $\smash{\targetEst}$ with $\smash{\weiRes^{1:M}}$ more homogeneous than $\smash{\Wei^{1:N}}$, in a suitable sense, whilst maintaining \emph{unbiasedness} i.e., that $\smash{\E[\targetEstRes(\testFun)|\calF] = \targetEst(\testFun)}$, for any suitably integrable test function $\testFun\colon \spaceX \to \reals$ where $\calF\coloneqq \sigma(\StaAll, \WeiAll)$ is the $\sigma$-algebra generated by $(\StaAll, \WeiAll)$. 

We will refer to the $\smash{\weiRes^m} \geq 0$ as post-resampling `weights', though it is possible that $\sum_{m=1}^M \weiRes^m \neq 1$ (equality holds for most of the commonly used resampling schemes and must hold at least in expectation, i.e., $\smash{\E[\sum_{m=1}^M \weiRes^m|\calF]} = 1$, almost surely, for resampling schemes that are unbiased). 

\begin{example}[sampling--importance resampling]
  Weighted samples $(\StaAll, \WeiAll)$ arise in \emph{\gls{SNIS}}. Here, we are interested in expectations of the form $\smash{\pi(\testFun) = \int_\spaceX \testFun(\sta) \pi(\diff \sta)}$, for a probability measure $\smash{\pi(\diff \sta)} \propto G(\sta) Q(\diff \sta)$ whose normalising constant is typically intractable; $Q$ is a probability measure on $\spaceX$, $G\colon \spaceX \to [0, \infty)$ is some non-negative $Q$-integrable function and $\smash{\testFun\colon \spaceX \to \reals}$ is some $\pi$-integrable function. \gls{SNIS} then estimates $\pi(\testFun)$ as $\smash{\targetEst(\testFun)} = \smash{\sum_{n=1}^N W^n \testFun(\Sta^n)}$, for $\smash{\Sta^1, \dotsc, \Sta^N \iidSim Q}$ and (self-normalised) importance weights $\smash{W^n \coloneqq G(\Sta^n) / \sum_{l=1}^N G(\Sta^l)}$. After resampling, expectations $\pi(\testFun)$ can again be estimated by $\smash{\targetEstRes(\testFun)}$. This is known as the \emph{\gls{SIR}} estimator \citep{rubin1987calculation}. The unbiasedness condition then implies that the \gls{SIR} estimate does not incur additional bias compared to the original \gls{SNIS} estimate. 
\end{example}

\subsubsection{Specific class of resampling schemes treated in this work}

Although resampling is very general and imposes almost no constraints on how $\smash{(\StaResAll, \weiResAll)}$ is produced, we focus in the sequel on the a class of resampling schemes which specify the resampled particles as $\smash{\StaRes^m \coloneqq \Sta^{\Anc^m}}$, where $\Anc^m$ is an \emph{ancestor index} taking a value in $[N]$. Definition~\ref{def:resampling_distribution} gives a sampling distribution for $\AncAll \coloneqq \Anc^{1:M}$ and for some associated optional collection of auxiliary variables $\Aux$ used in the simulation of $\AncAll$. The latter take values in some set $\spaceU$; if no such auxiliary variables are used, we set $\spaceU \coloneqq \emptyset$ and suppress $\Aux$ and $\aux$ from the notation. 

\begin{definition}[resampling distribution]\label{def:resampling_distribution}
  A distribution $\resDist(\ancAll, \diff \aux) = \Prob(\AncAll = \ancAll, \Aux \in \diff \aux)$ on $[N]^M \times \spaceU$ is called a a \emph{resampling distribution} if, for any $n \in [N]$, $\Wei^n > 0 \Longrightarrow \sum_{m=1}^M \Prob(\Anc^m = n) > 0$. 
\end{definition}

For any $m \in [M]$, we define the following notation: $\resDist^m(a, \diff \aux) \coloneqq \Prob(\Anc^m = a, \Aux \in \diff \aux)$ is the marginal distribution of $(\Anc^m\!, \Aux)$ under $\resDist$; $\resDist^m(a) \coloneqq \Prob(\Anc^m = a)$ is the marginal distribution of $\Anc^m$ under $\resDist$; $\resDist^{-m}(\ancAll^{-m}|a, \aux) \coloneqq \Prob(\AncAll^{-m} = \ancAll^{-m}| \Anc^m = a, \Aux = \aux)$ (assuming $M > 1$) is the full conditional distribution of $\AncAll^{-m} \coloneqq (A^1, \dotsc, A^{m-1}, \Anc^{m+1}\!, \dotsc, A^M)$ under $\resDist$.

\begin{example}[toy resampling scheme] \label{ex:toy:1}
 Assume $N = M = 2$ and $\spaceU = \emptyset$ and $\Wei^1 \in (0,1)$ so that $\Wei^2 = 1 - \Wei^1 \in (0, 1)$.  For some $\alpha, \beta \in [0,1]$, let $\resDist^1 = \alpha \delta_1 + (1-\alpha)\delta_2$ and $\resDist^2 = \beta \delta_1 + (1-\beta)\delta_2$ be the marginal distributions of $\resDist$. Then $\resDist$ is a resampling distribution if $0 < \alpha + \beta < 2$. Because then each particle is chosen as an ancestor with positive probability -- for any joint distribution $\resDist$ with the above-mentioned marginals.
 \end{example}

\begin{definition}[index distribution]
 Given $n \in [N]$, a conditional distribution $\indDist(m, \diff \aux|n)$ on $[M] \times \spaceU$ is called an \emph{index distribution}.
\end{definition}

The key r\^ole of index distributions for validity of (conditional) resampling is the focus of this work.

\begin{definition}[resampling scheme]\label{def:resampling_scheme}
  A \emph{resampling scheme} is a tuple $\resScheme$ consisting of (1) a resampling distribution $\resDist$ and (2) a family of index distributions $\indDistAll = (\indDist(\ccdot|n))_{n \in [N]}$, which generates $(\StaResAll, \weiResAll)$ as summarised in Algorithm~\ref{alg:resampling}, where
\begin{align}
   \weiResFun(m, n, \aux) \coloneqq 
   \frac{\Wei^n \indDist(m, \aux|n)}{M^{-1}\resDist^m(n, \aux)}. \label{eq:main_valid_weight_definition}
\end{align}
\end{definition}

\begin{figure}[htb]
\algorithmSepAbove
\begin{algorithm}[H]
\caption{Resampling}\label{alg:resampling}
\begin{algorithmic}[1]
  \State\label{enum:alg:resampling:sample_randomness}Sample $(\AncAll, \Aux) = (A^{1:M}, \Aux) \sim \resDist(\ancAll, \diff \aux)$.
  \State Set $\smash{\StaRes^m \coloneqq \Sta^{A^m}}$ and $\weiRes^m \coloneqq \weiResFun(m, \Anc^m\!, \Aux) / M$, for $m \in [M]$.
\label{enum:alg:resampling:sample_ancestors}
\end{algorithmic}
\end{algorithm}
\algorithmSepBelow
\end{figure}

When necessary, we will make the dependence of $\resDist$, $\resDist^m$, $\resDist^{-m}(\ccdot;n)$, $\indDist(\ccdot|n)$ and $\weiResFun$ on the weights $\WeiAll = W^{1:N}$ explicit in the notation by writing $\resDist = \resDist(\ccdot; \WeiAll)$ and $\resDist^m = \resDist^m(\ccdot; \WeiAll)$, $\resDist^{-m}(\ccdot|n; \WeiAll)$, $\indDist(\ccdot|n; \WeiAll)$ and $\weiResFun(\ccdot; \WeiAll)$; and, in principle, these could also depend on $\StaAll$. The dependence of these quantities (and also of $\spaceU$) on $N$ and $M$ will always be clear from the context.

\subsubsection{Sufficient condition for unbiasedness}

A frequently-imposed requirement on resampling schemes is that $\smash{\targetEstRes}$ is an unbiased estimate of $\targetEst$ in the following sense, where we remember that $\calF = \sigma(\StaAll, \WeiAll)$ is the $\sigma$-algebra generated by $(\StaAll, \WeiAll)$.

\begin{definition}[unbiasedness] \label{def:unbiasedness}
  A resampling scheme $\resScheme$ is \emph{unbiased} if 
  \begin{align}
  \E[\targetEstRes(\testFun) | \calF] = \targetEst(\testFun), 
  \end{align}
  almost surely, for any bounded continuous functions $\testFun\colon \spaceX \to \reals$.
\end{definition}

\begin{definition}[proper weighting] \label{def:proper_weighting}
   A resampling scheme $\resScheme$ is called \emph{properly weighted} if \eqref{eq:main_valid_weight_definition} is well defined, i.e., if for any $(m,n,\aux) \in [M] \times [N] \times \spaceU$, $\resDist^m(n, \aux) = 0$  $\Longrightarrow$ $\Wei^n \indDist(m, \aux|n) = 0$.
\end{definition}

\begin{example}[toy resampling scheme, continued] \label{ex:toy:2}
Consider the following families of index distributions $\indDistAll_i \coloneqq (\indDist_i(\ccdot|n))_{n \in [N]}$, for $i \in [2]$:
  \begin{enumerate}
     \item $\smash{\indDist_1(\ccdot|1) \propto \alpha \delta_1 +  \beta \delta_2}$; $\smash{\indDist_1(\ccdot|2) \propto (1- \alpha) \delta_1 + (1-\beta) \delta_2}$;
     \item $\smash{\indDist_2(\ccdot|1) = \indDist_2(\ccdot|2) \coloneqq \dUnif_{[2]}}$.
 \end{enumerate}
If $\alpha, \beta \in (0,1)$ then $\resSchemeNum{1}$ and $\resSchemeNum{2}$ are both properly weighted because the numerator and denominator in \eqref{eq:main_valid_weight_definition} is are strictly positive for any $(m, n) \in [2]^2$. If $\alpha = \beta = 1$ then $\resSchemeNum{1}$ is still properly weighted because $\resDist^m(n) = \indDist_1(m|n) = \ind\{m = n\}$ so that the denominator in \eqref{eq:main_valid_weight_definition} is zero only if the numerator is zero. However, in this case,
However, $\resSchemeNum{2}$ is not properly weighted because for $(m, n) \in [2]^2$ with $m \neq n$, the denominator in \eqref{eq:main_valid_weight_definition} is zero while the numerator remains strictly positive. Put differently, the second choice of index distribution allows for assignments incompatible with the resampling distribution causing $\weiResFun$ to be undefined (albeit on a set which is null w.r.t.\ $\resDist$).
\end{example}

The following is the main result of this section. It shows that proper weighting is sufficient for unbiasedness. Motivated by the fact that $\smash{\targetEst(f) = \E_{A \sim \dCat(\WeiAll)}[f(\Sta^A)|\calF]}$, the proof interprets resampling as importance sampling (with one sample point) targetting $\dCat(\WeiAll)$. More precisely, we perform importance sampling on the extended space that includes $(K, A, \AncAll, \Aux)$, where $K\in [M]$ is the index of a single ancestor index and $A \in [N]$ is its value which is almost surely equal to $A^K$. Proper weighting then guarantees that the target distribution of this importance-sampling construction  on $[M] \times [N] \times [N]^M \times \spaceU$: 
\begin{align}
 \mu(k, \anc, \ancAll, \diff \aux)
  & \coloneqq
  \smash{\dCat(\anc; \WeiAll) \indDist(k, \diff \aux|a) \delta_\anc(\anc^k) \resDist^{-k}(\ancAll^{-k}|\anc^k\!, \aux)}, 
  \label{eq:prop:proper_weighting_implies_unbiasedness:proof:1}
 \end{align}
(which admits $\dCat(\WeiAll)$ as a marginal) is absolutely continuous w.r.t.\ the sampling distribution
 \begin{align}
     q(k, a, \ancAll, \diff\aux)
     & \coloneqq \rho(\ancAll, \diff \aux)
     \dUnif_{[M]}(k) \delta_{a^k}(a).
     \label{eq:prop:proper_weighting_implies_unbiasedness:proof:2}
 \end{align}


\begin{proposition}
\label{prop:proper_weighting_implies_unbiasedness}
  A properly weighted resampling scheme is unbiased.
\end{proposition} 
\begin{namedProof}
 For the moment, assume that $M > 1$. 
 By  Definition~\ref{def:proper_weighting}, $\mu(\anc, k, \ancAll, \diff \aux) \ll q(k, a, \ancAll, \diff\aux)$, so that the following change of measure---with Radon--Nikodym derivative $[\diff \mu / \diff q](k,a,\ancAll,\aux) 
 = \weiResFun(k, a, \aux)$ given in \eqref{eq:main_valid_weight_definition}---is valid, where $\smash{\E_q[\ccdot|\calF]}$ is shorthand for $\smash{\E_{(K, A, \AncAll, \Aux)\sim q}[\ccdot|\calF]}$:
 \begin{align}
     \!\!\!\!\targetEst(f)
     = \sum_{n=1}^N W^n f(\Sta^n)
     & = \E_{A \sim \dCat(\WeiAll)}[f(\Sta^A)|\calF]
     = \E_{(K, A, \AncAll, \Aux)\sim \mu}[f(X^A)|\mathcal{F}]\\[-2ex]
    & = \E_q\biggl[ \frac{\diff \mu}{\diff q}(K, A, \AncAll, \Aux) f(X^A) \,\bigg|\,\mathcal{F}\biggr]\\
     & = \E_q\biggl[ \E_q\biggl(\frac{\diff \mu}{\diff q}(K, A, \AncAll, \Aux) f(X^A) \,\bigg|\, \sigma(\AncAll, \Aux) \vee \mathcal{F} \biggr) \,\bigg|\,\mathcal{F}\biggr]\\
     & = \E_q\biggl[ \frac{1}{M} \sum_{m=1}^M \weiResFun(m, A^m\!, \Aux)  f(X^{A^m}) \,\bigg|\, \mathcal{F}\biggr]
      = \E[\targetEstRes(f)|\mathcal{F}].
 \end{align}
 For $M = 1$, the proof is analogous except that we omit $\smash{\resDist^{-k}(\ancAll^{-k}|\anc^k\!, \aux)}$ in \eqref{eq:prop:proper_weighting_implies_unbiasedness:proof:1}.
 \hfill \qedwhite
\end{namedProof}

\subsubsection{Standard choice of index distribution}

The proper weighting definition leaves a degree of freedom in choosing the family of index distributions. The following class of resampling schemes provides a standard choice which automatically guarantees proper weighting by taking $\indDist(m, \diff \aux|n) \propto \resDist^m(n, \diff \aux)$.


\begin{definition}[balance] \label{def:balance}
  A resampling scheme $\resScheme$ is \emph{balanced} if
    for any $(m, n, \aux) \in [M] \times [N] \times \spaceU$:
    \begin{align}
    \Wei^n > 0 \quad \Longrightarrow \quad
   \indDist(m, \diff \aux|n) 
 =\frac{\resDist^m(n, \diff \aux)}{\sum_{l=1}^M \resDist^l(n)}. \label{eq:balanced_index_distribution}
\end{align}
\end{definition}

The name \emph{balanced} is motivated by the fact that \eqref{eq:balanced_index_distribution} is essentially the \emph{balance heuristic} for multiple importance sampling from \citet{veach1995optimally} (see also \citet{medina2019revisiting} for further discussions on the balance heuristic and its relationship with extended state-space justifications of importance sampling similar to \eqref{eq:prop:proper_weighting_implies_unbiasedness:proof:1}--\eqref{eq:prop:proper_weighting_implies_unbiasedness:proof:2}).

Within the importance-sampling construction used in the proof of Proposition~\ref{prop:proper_weighting_implies_unbiasedness}, \eqref{eq:balanced_index_distribution} implies that $\indDist(m, \diff \aux|n) = q(m, \diff \aux|n)$ 
(whenever this conditional distribution is well defined). Balanced resampling schemes have that $\weiResFun(m, n, u) = M W^n / \sum_{l=1}^M \resDist^l(n)$ is almost-surely constant in $(m, \aux)$, i.e., in the proof of Proposition~\ref{prop:proper_weighting_implies_unbiasedness}, they perform importance sampling on a smaller space which does not include $(K, \Aux)$. 

As Section~\ref{sec:examples} will demonstrate, most known resampling schemes are balanced, i.e., they implicitly use the index distribution from \eqref{eq:balanced_index_distribution}. The definition of proper weighting implies the following result. 

\begin{proposition} \label{prop:balance_implies_proper_weighting}
  A balanced resampling scheme is properly weighted.
\end{proposition}

The following example illustrates that using a balanced resampling scheme not only guarantees proper weighting but also ensures that the conditional variance of resampling is robust to single small marginal probabilities $\resDist^m(\Anc^m)$ under the resampling distribution.

\begin{example}[toy resampling scheme, continued] \label{ex:toy:3}
 Assume that $\alpha, \beta \in (0,1)$. Then as discussed above, $\resSchemeNum{1}$ and $\resSchemeNum{2}$ are both properly weighted. However, $\resSchemeNum{1}$ is balanced while $\resSchemeNum{2}$ is not (unless $\alpha = \beta = 1/2$). For simplicity. let $\testFun \equiv 1$ and fix $\beta, \Wei^1 \in (0,1)$. Then:
 \begin{enumerate}
     \item the balanced scheme $\resSchemeNum{1}$ has $\smash{\sup_{\alpha \in (0,1)} \var[\targetEstRes(\testFun)|\calF] < \infty}$;
     \item the imbalanced scheme $\resSchemeNum{2}$ has $\smash{\lim_{\alpha \downarrow 0} \var[\targetEstRes(\testFun)|\calF] = \lim_{\alpha \uparrow 1} \var[\targetEstRes(\testFun)|\calF] = \infty}$.
\end{enumerate}
Intuitively, $\indDistAll_1$ is `adapted' to the proposal distribution in the importance-sampling construction. For instance, as $\alpha \downarrow 0$, the probability $\resDist^1(1)$ vanishes. The family $\indDistAll_1$ accounts for this since $\alpha \downarrow 0$ implies $\indDist_1(1|1) \propto \resDist^1(1) \downarrow 0$. In contrast, $\indDistAll_2$ fixes $\indDist_2(1|1) = 1/2$. 
\end{example}

As noted above, a frequent goal of resampling is to make the resampled weights $\weiRes^{1:M}$ more homogeneous than $\Wei^{1:N}$. The following class of resampling schemes achieves $\weiRes^m = 1/M$, for all $m \in [M]$.

\begin{definition}[weight-homogeneous resampling scheme] \label{def:weight_homogeneity}
 A balanced resampling scheme $\resScheme$ is called \emph{weight-homogeneous} if $\weiResFun \equiv 1$ (almost surely), i.e., if for all $n \in [N]$:
 \begin{align}
    \sum_{m=1}^M \resDist^m(n) = MW^n. \label{eq:conventional_unbiasedness_condition}
\end{align}
\end{definition}

Since achieving $\weiRes^m = 1/M$ is frequently the goal of resampling, \eqref{eq:conventional_unbiasedness_condition} is itself considered \emph{the} unbiasedness condition \citep{crisan1998discrete}. We will thus call \eqref{eq:conventional_unbiasedness_condition} the conventional `unbiasedness' condition.

\begin{example}[toy resampling scheme, continued] \label{ex:toy:4}
 Simple algebra shows that the balanced resampling scheme $\resSchemeNum{1}$ is weight-homogeneous if $\alpha + \beta = 2 \Wei^1$.
\end{example}

\begin{definition}[persistence] \label{def:persistence}
 Let $\smash{B_+ \coloneqq
 \{ (\ancAll, \aux) \in [N]^N \times \spaceU | \sum_{m=1}^M \weiResFun(m, \anc^m\!, \aux) > 0 \}}$ be the event that the sum of the post-resampling weights under the  resampling scheme  $\resScheme$ is strictly positive. Then $\resScheme$ is called \emph{persistent} if $\smash{\resDist(B_+; W^{1:N})} = 1$,
  for any $\smash{W^1,\dotsc,W^N \geq 0}$ with $\smash{\sum_{n=1}^N W^n = 1}$.
\end{definition}

\begin{example}[toy resampling scheme, continued] \label{ex:toy:5}
 If $\alpha, \beta \in (0,1)$ then both $\resSchemeNum{1}$ and $\resSchemeNum{2}$ are persistent. However, the properly weighted resampling scheme $\resSchemeNum{3}$ given by $\indDist_3(\ccdot|n) = \delta_n$, for $n \in [2]$, is not because $\Anc^{1:2} = (2,1)$ and hence $\weiRes^1 = \weiRes^2 = 0$ has positive probability under $\resSchemeNum{3}$
\end{example}

Persistence guarantees that the post-resampling weights $\smash{\weiRes^m}$ can (almost surely) be self-normalised so that another round of resampling could be applied to $\smash{(\StaRes^{1:M}, \WeiRes^{1:M})}$, where $\smash{\WeiRes^m \coloneqq \weiRes^m / \sum_{l=1}^M \weiRes^l}$. The following sufficient condition for persistence is immediate from the definition.

\begin{proposition}\label{prop:sufficient_condition_for_persistence}
  A balanced resampling scheme $\resScheme$ is persistent if for all $n \in [N]$: $\sum_{m=1}^M \resDist^m(n) > 0 \Longrightarrow W^n > 0$.
\end{proposition}

\begin{remark}\label{lab:sufficient_conditions_for_persistence}
  The sufficient condition for persistence given in Proposition~\ref{prop:sufficient_condition_for_persistence} holds immediately if the resampling scheme $\resScheme$ is weight-homogeneous.
\end{remark}
%

\subsection{Conditional resampling}
We will now turn to \emph{conditional} resampling.  This is a key ingredient in  \gls{CSMC} methods but we present it here without reference to the latter (the extension to \gls{CSMC} algorithms will be discussed in Section~\ref{subsec:csmc}). Admittedly, this makes our discussion below somewhat abstract. However, we believe that abstraction is justified by the fact that doing so avoids the notational burden incurred by \gls{CSMC} methods.

Throughout this section, we assume that $M \geq 2$ and that the resampling scheme $\resScheme$ is persistent (so that the post-resampling `weights' can be normalised). Given some $a \in [N]$ with $\smash{W^a > 0}$, conditional resampling (Algorithm~\ref{alg:conditional_resampling}) forms a (possibly random) probability measure
\begin{align}
  \targetEstRes^a \coloneqq 
  \sum_{m=1}^M \frac{\weiRes^m}{\sum_{l=1}^M \weiRes^l} \delta_{\StaRes^m},
\end{align}
in such a way that the support of $\targetEstRes^a$ includes $\Sta^a$. 
\begin{figure}[htb]
\algorithmSepAbove
\begin{algorithm}[H]
\caption{conditional resampling}\label{alg:conditional_resampling}
\begin{algorithmic}[1]
  \Require
  $a \in [N]$ with $\smash{W^a > 0}$.
  \State Sample $(K, \Aux, \AncAll) \sim \indDist(k, \aux |a) \delta_{a}(a^k) \resDist^{-k}(\ancAll^{-k}|\anc^k\!, \aux)$. \label{enum:alg:conditional_resampling:sample_k}
  \State Set $\smash{\StaRes^m \coloneqq \Sta^{A^m}}$ as well as $\weiRes^m \coloneqq \weiResFun(m, \Anc^m\!, \Aux) / M$, for $m \in [M]$.
\end{algorithmic}
\end{algorithm}
\algorithmSepBelow
\end{figure}

The following requirement says that if we first sample $A = a$ with probability $W^a$ and then perform conditional resampling, $\smash{\targetEstRes^a}$ should again induce a random measure that is unbiased for $\targetEst$.

\begin{definition}[invariance] \label{def:invariance}
 A persistent resampling scheme $\resScheme$ is \emph{invariant} if \[\smash{\sum_{a = 1}^N \Wei^a \E[\targetEstRes^a(\testFun)| \calF ]
   = \targetEst(\testFun)},\] almost surely, for all bounded continuous functions $\testFun\colon \spaceX \to \reals$.
\end{definition} 

The terminology `invariance' is motivated by the fact that Definition~\ref{def:invariance} holds if executing Algorithm~\ref{alg:conditional_resampling} and then returning the ancestor index $\Anc^L$ with probability proportional to $\weiRes^L$ induces a $\dCat(\WeiAll)$-invariant Markov kernel. The main result of this section is the following.

\begin{proposition}[proper weighting implies invariance] \label{prop:proper_weighting_implies_invariance}
   A properly weighted and persistent resampling scheme is invariant.
\end{proposition}
\begin{namedProof} 
 We extend the distribution $\mu$ from the proof of Proposition~\ref{prop:proper_weighting_implies_unbiasedness}  so that it represents the law of first sampling $A \sim \dCat(\WeiAll)$, then running Algorithm~\ref{alg:conditional_resampling} (with $A = a$) and finally selecting the $L$th ancestor index with probability proportional to the $L$th post-resampling `weight' (and calling it $B$).
 \begin{align}
  \mu(k, a, \ancAll, \diff \aux, b, l) \coloneqq \dCat(a; \WeiAll) \indDist(k, \diff \aux |a) \delta_a(a^k) \resDist^{-k}(\ancAll^{-k}|\anc^k\!, \aux) \frac{\weiResFun(l, \anc^l\!, \aux)}{\sum_{m=1}^M \weiResFun(m, \anc^m\!, \aux)} \delta_{a^l}(b).
 \end{align}
 By persistence and proper weighting, this distribution is well defined. Simple algebra verifies the symmetry $\mu(k, a, \ancAll, \diff \aux, b, l) = \mu(l, b, \ancAll, \diff \aux, a, k)$ which in turn implies
 \begin{align}
     \sum_{a = 1}^N \Wei^a \E[\targetEstRes^a(\testFun)| \calF ] = \E_\mu[f(X^{B})|\calF] = \E_\mu[f(X^{A})|\calF] = \targetEst(f),
 \end{align}
 where $\smash{\E_\mu[\ccdot|\calF]}$ denotes expectation w.r.t.\ $(K, A, \AncAll, \Aux, B, L) \sim \mu$. 
 \hfill \qedwhite
\end{namedProof}

\subsection{Implementation of conditional resampling with fixed reference index}

In practice, conditional resampling (within \gls{CSMC} algorithms) is often implemented by placing the ancestor index $a$ in a fixed location $K = 1$ or $K = M$, say, in Line~\ref{enum:alg:conditional_resampling:sample_k} of Algorithm~\ref{alg:conditional_resampling}. While this strategy is generally invalid (invariance is lost), the next proposition (proved in Appendix~\ref{app:sec:proofs_for_section_2}) shows that it \emph{is} valid when the resampling distribution is symmetric in the following sense.

\begin{definition}[reindexability] \label{def:symmetry} 
A resampling scheme $\resScheme$ is called \emph{reindexable} if $\smash{\E_{k,a}[f(\AncAll)|\calF]} \allowbreak=\allowbreak \smash{\E_{1,a}[f(\AncAll)|\calF]}$, for any $a \in [N]$ with $\smash{\Wei^a > 0}$, any $k \in [M]$ and any function $\smash{f\colon [N]^M \to \reals}$ which is symmetric in its $M$ arguments, where $\smash{\E_{k,a}}$ denotes expectation w.r.t.\ $\smash{\AncAll \sim \delta_{a}(a^k) \resDist^{-k}(\ancAll^{-k}|\anc^k, \aux)}$.
\end{definition}

\begin{proposition}\label{prop:alternative_lambda_in_exchangeable_resampling_schemes}
  Let $\resScheme$ be a balanced and persistent resampling scheme. If $\resScheme$ is reindexable then invariance still holds if Line~\ref{enum:alg:conditional_resampling:sample_k} of Algorithm~\ref{alg:conditional_resampling} is replaced by
\begin{enumerate}
\item[\ref{enum:alg:conditional_resampling:sample_k}':]
Sample $(K, \Aux, \AncAll) \sim \indDist'(k|a) \resDist^k(\aux |a) \delta_{a}(a^k) \resDist^{-k}(\ancAll^{-k}|\anc^k\!, \aux)$,
\end{enumerate}
 where $\indDist'(\ccdot|n)$ is an arbitrary distribution on $[M]$, for any $n \in [N]$.
\end{proposition}

The reindexability assumption of Proposition~\ref{prop:alternative_lambda_in_exchangeable_resampling_schemes} is satisfied for exchangeable resampling schemes.


\begin{definition}[exchangeability]
 A resampling scheme $\resScheme$ is \emph{exchangeable} if the marginal distribution $\resDist(\ancAll)$ is exchangeable. 
\end{definition}

\begin{proposition}
 A resampling scheme is reindexable if it is exchangeable.
\end{proposition}

\begin{remark}\label{rem:random_permutation_and_shift}
For non-exchangeable resampling schemes, reindexability can be satisfied by permuting or shifting the order of the ancestor indices uniformly at random. Our work shows that such random permutations/shifts are only necessary if one wishes to implement conditional resampling in such a way that $K$ is always set to $1$ or $M$, say, in Line~\ref{enum:alg:conditional_resampling:sample_k} of Algorithm~\ref{alg:conditional_resampling}. Otherwise, they only add computational cost. More details are given in Appendix~\ref{app:subsec:random_permutation}.
\end{remark}

\section{Examples}
\label{sec:examples}

We now demonstrate that a number of well known resampling schemes $\resScheme$ can be viewed as special cases of our framework. In particular, this allows us to present valid \emph{conditional} versions of popular resampling schemes that are not exchangeable, without the need for randomly permuting or shifting the order of the generated ancestor indices.

We start with a number of \emph{elementary} resampling schemes (multinomial, stratified, systematic or killing resampling and even `trivial' resampling, i.e., not resampling). Then, we discuss \emph{composite} resampling schemes which build more sophisticated resampling schemes from elementary ones.

\begin{remark}\label{rem:integrating_out_auxiliary_variables_in_balanced_resampling_schemes}
  For balanced resampling schemes, the auxiliary variables $\Aux$ can be integrated out without changing the law of $(\StaRes^{1:M}, \weiRes^{1:M})$ conditional on $\calF$. Hence, we sometimes include $\Aux$ in the balanced resampling schemes below purely to simplify the presentation. But we could equally work with $\spaceU = \emptyset$.
\end{remark}

\subsection{Elementary resampling schemes}
\label{subsec:elementary_resampling_schemes}

The popular elementary resampling schemes discussed in this section are persistent and balanced and all (except `trivial' resampling) are even weight-homogeneous.

\subsubsection{Multinomial resampling}
\label{subsubsec:multinomial_resampling}

\emph{Multinomial} resampling \citep{rubin1987calculation}---also known as \emph{stochastic sampling with replacement} in the genetic algorithms literature \citep{baker1987reducing}---takes $\spaceU \coloneqq \emptyset$ and $\resDist(\ancAll) \coloneqq \smash{\prod_{m=1}^M \resDist^m(a^m)}$, where $\smash{\resDist^m \coloneqq \dCat(\WeiAll)}$, and $\smash{\indDist(m|n) \coloneqq \dUnif_{[M]}(m)}$ and hence $\weiResFun \equiv 1$. Clearly, multinomial resampling is weight-homogeneous (and hence balanced and persistent) and also exchangeable (so that Proposition~\ref{prop:alternative_lambda_in_exchangeable_resampling_schemes} would allow us to simply set $K \coloneqq 1$, say, in Algorithm~\ref{alg:conditional_multinomial_resampling}). Line~\ref{alg:multinomial_resampling:2} of Algorithm~\ref{alg:multinomial_resampling} and Line~\ref{alg:conditional_multinomial_resampling:2} of Algorithm~\ref{alg:conditional_multinomial_resampling} can be implemented in $\bo(\max\{M, N\})$ operations using \emph{aliasing} methods \citep{walker1974new, walker1977efficient}.

\begin{figure}[htb]
\algorithmSepAbove
\begin{algorithm}[H]
\caption{multinomial resampling}\label{alg:multinomial_resampling}
\begin{algorithmic}[1]
   \State Sample $\smash{A^m \sim \dCat(\WeiAll)}$, for $m \in [M]$.
   \State Set $\smash{\StaRes^m \coloneqq \Sta^{A^m}}$ and $\smash{\weiRes^m \coloneqq 1/M}$, for $m \in [M]$. \label{alg:multinomial_resampling:2}
\end{algorithmic}
\end{algorithm}
\algorithmSepBelow
\end{figure}

\begin{figure}[htb]
\algorithmSepAboveAlt
\begin{algorithm}[H]
\caption{conditional multinomial resampling}\label{alg:conditional_multinomial_resampling}
\begin{algorithmic}[1]
  \Require $a \in [N]$ with $\smash{W^a > 0}$.
   \State Sample $\smash{K \sim \dUnif_{[N]}}$ and set $\smash{A^K \coloneqq a}$.
   \State Sample $\smash{A^m \sim \dCat(\WeiAll)}$, for $m \in [M] \setminus \{K\}$. \label{alg:conditional_multinomial_resampling:2}
   \State Set $\smash{\StaRes^m \coloneqq \Sta^{A^m}}$ and $\smash{\weiRes^m \coloneqq 1/M}$, for $m \in [M]$.
\end{algorithmic}
\end{algorithm}
\algorithmSepBelow
\end{figure}

\subsubsection{Stratified resampling}
\label{subsubsec:stratified_resampling}

\emph{Stratified} resampling \citep{kitagawa1996monte} takes $\spaceU \coloneqq [0,1]^M$ and, writing $\Aux = \Aux^{1:M}$:
\begin{align}
  \resDist(\ancAll, \diff \aux) 
  & \coloneqq \prod_{m=1}^M \dUnif_{[0,1]}(\diff \aux^m) \delta_{F^{-1}([u^m + m - 1] / M )}(\anc^m),
\end{align}
where $F(n) \coloneqq \sum_{l=1}^n \Wei^l$, for $n \in [N]$ and its generalized invers $F^{-1}(p) \coloneqq \min\{n \in [N] \mid p \leq F(n)\}$, for $p \in [0,1]$. For what follows, it will be convenient to use that for any $u \in [0, 1]$:
\begin{align}
  n = F^{-1}\biggl(\frac{u+m-1}{M}\biggr) 
  & \Longleftrightarrow u \in I^m(n),
\end{align}
for any $(m, n, u) \in [M] \times [N]  \times [0,1]$, where $\{I^m(n)\}_{n \in [N]}$ is the partition of $(0, 1]$ given by
\begin{align}
  I^m(n) 
  & \coloneqq 
  (\max\{0, M F(n-1) -m +1\}, \min\{1, M F(n) - m +1\}].
\end{align}
with convention $F(0) \coloneqq 0$. With this notation, we can express the resampling distribution equivalently as
  $\smash{\resDist(\ancAll, \diff \aux)
  \coloneqq \prod_{m=1}^M \resDist^m(a^m) \dUnif_{I^m(a^m)}(\diff \aux^m)}$,
where 
\begin{align}
  \resDist^m(n) = \dCat(n; (\vol(I^m(l)))_{l \in [N]}). \label{eq:stratified_resampling_marginal}
\end{align}
Stratified resampling is not exchangeable. However, $\resDist$ satisfies the conventional `unbiasedness' condition \eqref{eq:conventional_unbiasedness_condition}: 
\begin{align}
  \sum_{m=1}^M \resDist^m(n)
= \sum_{m=1}^M \vol((\max\{m-1, M F(n-1)\}, \min\{m, M F(n)\}])
  = M \Wei^n.
\end{align}
Consequently, stratified resampling is weight-homogeneous (and hence balanced and thus properly weighted as well as persistent) with
\begin{align}
  \indDist(m, \aux|n) 
  & \coloneqq \frac{\rho^m(n, u)}{M W^n}\\
  & = \dCat(m; (\vol(I^l(n)) / (MW^n))_{l \in [M]}) \dUnif_{I^m(n)}(\aux^m) \smash{\prodSubstackAligned{l}{=}{1}{l}{\neq}{m}^{\smash{M}}} \dUnif_{[0,1]}(\aux^l). \label{eq:lambda_stratified_resampling} 
\end{align}

\begin{figure}[htb]
\algorithmSepAbove
\begin{algorithm}[H]
\caption{stratified resampling}\label{alg:stratified_resampling}
\begin{algorithmic}[1]
    \State Sample $\smash{U^m \sim \dUnif_{[0,1]}}$, for $m \in [M]$.
    \State Set $\smash{A^m \coloneqq F^{-1}([U^m + m - 1] / M)}$, for $m \in [M]$, via Algorithm~\ref{alg:inverse_cdf_sampling}. \label{alg:stratified_resampling:2}
    \State Set $\smash{\StaRes^m \coloneqq \Sta^{A^m}}$ and $\smash{\weiRes^m \coloneqq 1/M}$, for $m \in [M]$.
\end{algorithmic}
\end{algorithm}
\algorithmSepBelow
\end{figure}

\begin{figure}[htb]
\algorithmSepAboveAlt
\begin{algorithm}[H]
\caption{conditional stratified resampling}\label{alg:conditional_stratified_resampling}
\begin{algorithmic}[1]
  \Require $a \in [N]$ with $\smash{W^a > 0}$.
   \State Sample $\smash{K \sim \dCat((\vol(I^m(a)) / (MW^a))_{m \in [M]})}$.
   \State Sample $\smash{U^K \sim \dUnif_{I^K(a)}}$, and $\smash{U^m \sim \dUnif_{[0,1]}}$ for $m \in [M] \setminus \{K\}$. 
 \State Set $\smash{A^m \coloneqq F^{-1}([U^m + m - 1] / M)}$, for $m \in [M]$, via Algorithm~\ref{alg:inverse_cdf_sampling}. \label{alg:conditional_stratified_resampling:2}
  \State Set $\smash{\StaRes^m \coloneqq \Sta^{A^m}}$ and $\smash{\weiRes^m \coloneqq 1/M}$, for $m \in [M]$.
\end{algorithmic}
\end{algorithm}
\algorithmSepBelow
\end{figure}

Note that in Algorithm~\ref{alg:conditional_stratified_resampling}, by construction, $\smash{\Anc^K = a}$ with probability $1$. It is well known that Line~\ref{alg:stratified_resampling:2} in Algorithm~\ref{alg:stratified_resampling} and Line~\ref{alg:conditional_stratified_resampling:2} in 
Algorithm~\ref{alg:conditional_stratified_resampling} can be implemented in $\bo(\max\{M, N\})$ operations as outlined in Algorithm~\ref{alg:inverse_cdf_sampling}.

\begin{figure}[htb]
\algorithmSepAbove
\begin{algorithm}[H]
\caption{efficient stratified inverse transform method}\label{alg:inverse_cdf_sampling}
\begin{algorithmic}[1]
   \Require $\Wei^{1:N} \in [0,1]^N$ with $\sum_{n=1}^N \Wei^n = 1$ and $U^{1:M} \in [0,1]^M$.
   \State Set $Q^n \coloneqq \sum_{l=1}^n W^l$, for $n \in [N]$, and $T^m \coloneqq (U^m + m - 1) / M$, for $m \in [M]$.
   \State Set $i \coloneqq 1$, $j \coloneqq 1$.
   \While{$j \leq M$}
    \If{$T^j \leq Q^j$} set $A^j \coloneqq i$ and $j \leftarrow j + 1$.
    \Else{} Set $i \leftarrow i + 1$.
    \EndIf
  \EndWhile
  \State Set $\smash{\StaRes^m \coloneqq \Sta^{A^m}}$ and $\smash{\weiRes^m \coloneqq 1/M}$, for $m \in [M]$.
\end{algorithmic}
\end{algorithm}
\algorithmSepBelow
\end{figure}

\subsubsection{Systematic resampling}
\label{subsubsec:systematic_resampling}

\emph{Systematic} resampling \citep{carpenter1999improved}---also known as \emph{stochastic universal sampling} in genetic algorithms  \citep{baker1987reducing}---differs from stratified resampling in that it shares the same uniform random variable $U$ across all strata: 
\begin{align}
  \resDist(\ancAll, \aux) 
   \coloneqq  \dUnif_{[0,1]}(\aux) \prod_{m=1}^M \delta_{F^{-1}([u + m - 1] / M )}(\anc^m),
\end{align}
with $\spaceU \coloneqq [0,1]$.
Thus, $\smash{\resDist^m(n) = \vol(I^m(n))}$ as in stratified resampling \eqref{eq:stratified_resampling_marginal} so that systematic resampling again satisfies the conventional `unbiasedness' condition \eqref{eq:conventional_unbiasedness_condition}; and setting
\begin{align}
  \indDist(m, \aux|n) \coloneqq \frac{\rho^m(n, \aux)}{M W^n} = \dCat(m; (\vol(I^l(n)) / (MW^n))_{l \in [M]}) \dUnif_{I^m(n)}(\aux), \label{eq:lambda_systematic_resampling}
\end{align}
ensures that systematic resampling is weight-homogeneous (and hence balanced and thus properly weighted as well as persistent).

\begin{figure}[htb]
\algorithmSepAbove
\begin{algorithm}[H]
\caption{systematic resampling}\label{alg:systematic_resampling}
\begin{algorithmic}[1]
    \State Sample $\smash{U \sim \dUnif_{[0,1]}}$.u
    \State Set $\smash{A^m \coloneqq F^{-1}([U + m - 1] / M)}$, for $m \in [M]$ via Alg.~\ref{alg:inverse_cdf_sampling} with $\smash{U^1 = \ldots = U^M = U}$. \label{alg:systematic_resampling:2}
    \State Set $\smash{\StaRes^m \coloneqq \Sta^{A^m}}$ and $\smash{\weiRes^m \coloneqq 1/M}$, for $m \in [M]$.
\end{algorithmic}
\end{algorithm}
\algorithmSepBelow
\end{figure}

\begin{figure}[htb]
\algorithmSepAboveAlt
\begin{algorithm}[H]
\caption{conditional systematic resampling}\label{alg:conditional_systematic_resampling}
\begin{algorithmic}[1]
  \Require $a \in [N]$ with $\smash{W^a > 0}$.
    \State Sample $\smash{K \sim \dCat((\vol(I^m(a)) / (MW^a))_{m \in [M]})}$ and $\smash{U \sim \dUnif_{I^K(a)}}$. 
 \State Set $A^m \coloneqq F^{-1}([U + m - 1] / M )$, for $m \in [M]$ via Alg.~\ref{alg:inverse_cdf_sampling} with $U^1 = \ldots = U^M = U$. \label{alg:conditional_systematic_resampling:2}
  \State Set $\smash{\StaRes^m \coloneqq \Sta^{A^m}}$ and $\smash{\weiRes^m \coloneqq 1/M}$, for $m \in [M]$.
\end{algorithmic}
\end{algorithm}
\algorithmSepBelow
\end{figure}

Previously, \citet[][Algorithm~4]{chopin2015particle} derived a version of conditional systematic resampling which coincides with Algorithm~\ref{alg:conditional_systematic_resampling} up to an additional random shift (see Remark~\ref{rem:random_permutation_and_shift}).

\subsubsection{Killing resampling}
\label{subsubsec:killing_resampling}

Let $M = N$. \emph{Killing} resampling \citep{chopin2022resampling} \citep[see also][Section~2.5.3, for a closely related scheme]{delmoral2004feynman} takes $\spaceU \coloneqq \emptyset$, $\smash{\resDist(\ancAll)} \coloneqq \smash{\prod_{m=1}^M \resDist^m(a^m)}$, where, writing $\smash{W^\star} \coloneqq \smash{\max \{W^1, \dotsc, W^N\}}$,
\begin{align}
  \resDist^m(n) 
  & \coloneqq \frac{W^m}{W^\star} \delta_m(n)  + \Bigl(1 -  \frac{W^m}{W^\star}\Bigr) \dCat(n; \WeiAll). 
\end{align}
It can be verified that $\resDist$ thus satisfies the conventional `unbiasedness' condition \eqref{eq:conventional_unbiasedness_condition}. Thus, taking
\begin{align}
  \indDist(m|n)
  & \coloneqq \frac{\resDist^m(n)}{M W^n}
  =  \frac{W^\star - W^n +1}{MW^\star} \ind\{m = n\} + \frac{W^\star - W^m}{MW^\star} \ind\{m \in [M] \setminus \{n\}\}, 
  \label{eq:killing_resampling_lambda} 
\end{align}
means that killing resampling is weight-homogeneous (and hence properly weighted and persistent).

\begin{figure}[htb]
\algorithmSepAbove
\begin{algorithm}[H]
\caption{killing resampling}\label{alg:killing_resampling}
\begin{algorithmic}[1]
    \For{$m \in [M]$}
        \State \textbf{with probability} $\smash{W^m/W^\star}$ set $\smash{A^m \coloneqq m}$.
        \State \textbf{otherwise} sample $\smash{A^m \sim \dCat(\WeiAll)}$.
    \EndFor
    \State Set $\smash{\StaRes^m \coloneqq \Sta^{A^m}}$ and $\smash{\weiRes^m \coloneqq 1/M}$, for $m \in [M]$.
\end{algorithmic}
\end{algorithm}
\algorithmSepBelow
\end{figure}

\begin{figure}[htb]
\algorithmSepAboveSmall
\begin{algorithm}[H]
\caption{conditional killing resampling}\label{alg:conditional_killing_resampling}
\begin{algorithmic}[1]
  \Require $a \in [N]$ with $\smash{W^a > 0}$.
  \State Sample $\smash{K \sim \indDist(\ccdot|a)}$ defined in \eqref{eq:killing_resampling_lambda} and set $\smash{A^K \coloneqq a}$.
  \For{$m \in [M] \setminus\{K\}$}
        \State \textbf{with probability} $\smash{W^m/W^\star}$ set $\smash{A^m \coloneqq m}$.
        \State \textbf{otherwise} sample $\smash{A^m \sim \dCat(\WeiAll)}$.
  \EndFor
  \State Set $\smash{\StaRes^m \coloneqq \Sta^{A^m}}$ and $\smash{\weiRes^m \coloneqq 1/M}$, for $m \in [M]$.
\end{algorithmic}
\end{algorithm}
\algorithmSepBelow
\end{figure}

Previously, \citet{karppinen2023conditional} derived a version of conditional killing resampling which coincides with Algorithm~\ref{alg:conditional_killing_resampling} up to an additional random shift (see Remark~\ref{rem:random_permutation_and_shift}).

\subsubsection{Trivial resampling (i.e., not resampling)}
\label{subsubsec:trivial_resampling}

We now show that `not resampling' can be viewed as a resampling scheme (termed \emph{trivial} resampling) within our framework. This will become useful in Section~\ref{subsubsec:adaptive_resampling} where justify adaptive resampling as mixing trivial resampling with another, `non-trivial' resampling scheme.

Let $M \coloneqq N$. Then trivial resampling can be formalised as $\spaceU \coloneqq \emptyset$, $\resDist(\ancAll) \coloneqq \smash{\prod_{m=1}^M \resDist^m(a^m)}$, 
where $\smash{\resDist^m \coloneqq \delta_{m}}$. With $\smash{\indDist(\ccdot|n) \coloneqq \delta_n}$, trivial resampling is then balanced (and hence properly weighted) and persistent, and has $\smash{\weiResFun(m, n, \aux)} = \smash{M \Wei^n \ind\{m = n\}}$.

\begin{figure}[htb]
\algorithmSepAbove
\begin{algorithm}[H]
\caption{trivial resampling}\label{alg:trivial_resampling}
\begin{algorithmic}[1]
    \State Set $\smash{A^m \coloneqq m}$, for $m \in [M]$.
    \State Set $\smash{\StaRes^m \coloneqq \Sta^{m}}$ and $\smash{\weiRes^m \coloneqq W^m}$, for $m \in [M]$.
\end{algorithmic}
\end{algorithm}
\algorithmSepBelow
\end{figure}

\begin{figure}[htb]
\algorithmSepAboveAlt
\begin{algorithm}[H]
\caption{conditional trivial resampling}\label{alg:conditional_trivial_resampling}
\begin{algorithmic}[1]
  \Require $a \in [N]$ with $\smash{W^a > 0}$.
    \State Set $K \coloneqq a$ and $\smash{A^K \coloneqq a}$.
    \State Set $\smash{A^m \coloneqq m}$, for $m \in [M] \setminus \{K\}$.
    \State Set $\smash{\StaRes^m \coloneqq \Sta^{m}}$ and $\smash{\weiRes^m \coloneqq W^m}$, for $m \in [M]$.
\end{algorithmic}
\end{algorithm}
\algorithmSepBelow
\end{figure}

\subsection{Composite resampling schemes}
\label{subsec:composite_resampling_schemes}

Let $\resSchemeAlt$ denote some properly weighted and persistent resampling scheme (e.g., one of the elementary resampling schemes from the previous section). In this section, we illustrate that $\resSchemeAlt$ can be modified to obtain a new properly weighted and persistent resampling scheme $\resScheme$ with potentially more favourable properties. This permits a simple justification of many sophisticated resampling schemes that have been proposed in the literature such as residual resampling, resampling with weight re-ordering, adaptive resampling, optimal finite-state resampling and chopthin resampling. In particular, our framework shows how to implement conditional resampling for each of these.

\subsubsection{Residual resampling}
\label{subsubsec:residual_resampling}

Let $\smash{M_0 \coloneqq \sum_{n=1}^N \lfloor M W^n \rfloor}$ and
\begin{align}
  \smash{\alpha = \alpha^{1:M_0} = (\overbrace{1, \dotsc, 1}^{\mathclap{\text{$\lfloor M W^1 \rfloor$}\atop\text{times}}}, \overbrace{2, \dotsc, 2}^{\mathclap{\text{$\lfloor M W^2 \rfloor$}\atop\text{times}}}, \dotsc, \overbrace{N, \dotsc, N}^{\mathclap{\text{$\lfloor M W^N \rfloor$}\atop\text{times}}})}.
\end{align}
Additionally, let $\resSchemeAlt$ be some weight-homogeneous resampling scheme (e.g.\ any of the elementary resampling schemes from Section~\ref{subsec:elementary_resampling_schemes} except for `trivial' resampling) which generates $M - M_0$ ancestor indices on $[N]$ based on weights $\mathbf{V} \coloneqq V^{1:N}$, where
\begin{align}
  V^n
  & \coloneqq \frac{M W^n - \lfloor M W^n \rfloor}{M - M_0}.
\end{align}
\emph{Residual} resampling then takes the following resampling distribution (other orderings of $\smash{\ancAll = \anc^{1:M}}$ are possible and can facilitate `in place' implementations): 
\begin{align}
  \resDist(\ancAll, \aux) \coloneqq \delta_{\alpha}(a^{1:M_0}) \smashoperator{\sum_{\mathbf{b} \in [N]^{M-M_0}}}\resDistAlt(\mathbf{b}, \aux; \mathbf{V}) \delta_{\mathbf{b}}(a^{(M_0+1):M}).
\end{align}
Consequently,
  $\smash{\resDist^m(n)  = \ind\{m \in [M_0]\} \delta_{\alpha^m}(n) + \ind\{m \in [M] \setminus [M_0]\} \resDistAlt^{m-M_0}(n; \mathbf{V})}$.
It can be verified that $\resDist$ satisfies the conventional `unbiasedness' condition \eqref{eq:conventional_unbiasedness_condition}. Thus, setting
\begin{align}
  \indDist(m, \aux|n)
  & = \frac{\resDist^m(n, \aux)}{M W^n}  = p^n \dUnif_{\{l \in [M_0] \mid \alpha^l = n\}}(m) \resDistAlt(\aux) +
  (1 - p^n) \kappa(m-M_0, \aux|n; \mathbf{V}),
\end{align}
where $\smash{p^n \coloneqq \lfloor M W^n \rfloor / (MW^n)}$, makes the residual resampling scheme $\resScheme$ is weight-homogenous.

If $\resSchemeAlt$ is multinomial resampling then $\resScheme$ is called \emph{residual-multinomial resampling} \citep{liu1998sequential}---also called \emph{remainder stochastic sampling with replacement} in genetic algorithms \citep{baker1987reducing}---but other choices, e.g., \emph{residual-systematic resampling} \citep{higuchi1995resampling} are possible.

\begin{figure}[htb]
\algorithmSepAbove
\begin{algorithm}[H]
\caption{residual resampling}\label{alg:residual_resampling}
\begin{algorithmic}[1]
    \State Sample $(\mathbf{B}, \Aux) \coloneqq (B^{1:(M-M_0)}, \Aux) \sim \resDistAlt(\ccdot; \mathbf{V})$.
    \State Set $\AncAll \coloneqq (\alpha, \mathbf{B})$.
    \State Set $\smash{\StaRes^m \coloneqq \Sta^{A^m}}$ and $\smash{\weiRes^m \coloneqq 1/M}$, for $m \in [M]$.
\end{algorithmic}
\end{algorithm}
\algorithmSepBelow
\end{figure}

\begin{figure}[htb]
\algorithmSepAboveAlt
\begin{algorithm}[H]
\caption{conditional residual resampling}\label{alg:conditional_residual_resampling}
\begin{algorithmic}[1]
  \Require $a \in [N]$ with $\smash{W^a > 0}$.
  \State Sample $V \sim \dUnif_{[0,1]}$.
  \If{$V \leq \lfloor M W^a \rfloor /(M W^a)$}
    \State Sample $K \sim \dUnif_{\{m \in [M_0] \mid \alpha^m = a\}}$.
    \State Sample $(\AncAll, \Aux)$ via Algorithm~\ref{alg:residual_resampling}.
  \Else 
    \State Sample $(L, \mathbf{B}, \Aux) \sim \kappa(l, \aux |a; \mathbf{V}) \delta_{a}(b^l) \resDistAlt^{-l}(\mathbf{b}^{-l}|b^l, \aux; \mathbf{V})$.
     \State $K \coloneqq M_0 + L$ and $\AncAll \coloneqq (\alpha, \mathbf{B})$.
  \EndIf
  \State Set $\smash{\StaRes^m \coloneqq \Sta^{A^m}}$ and $\smash{\weiRes^m \coloneqq 1/M}$, for $m \in [M]$.
\end{algorithmic}
\end{algorithm}
\algorithmSepBelow
\end{figure}

Previously, \citet[][Algorithm~2]{chopin2015particle} derived a version of conditional residual-multinomial resampling which coincides with Algorithm~\ref{alg:conditional_residual_resampling} up to an additional random shift (see Remark~\ref{rem:random_permutation_and_shift}).

\subsubsection{Re-ordering weights prior to resampling}
\label{subsubsec:weight_re-ordering}

Let $\resSchemeAlt$ be a balanced and persistent resampling scheme. If $\resSchemeAlt$ is non-exchangeable, then it can sometimes be desirable to re-order the weights before resampling, e.g., to improve convergence rates \citep{kitagawa1996monte, douc2005comparison, gerber2019negative}.

More formally, let $\varpi \in \permutations_N$ be some 
$\calF$-measurable permutation (we only consider deterministic weight reordering for simplicity but random reorderings would be possible, too). Furthermore, write $\smash{\WeiAll^\varpi} = \smash{(W^{\varpi(1)}, \dotsc, W^{\varpi(N)})}$. We also now include the (order of the) weights explicitly into the resampling distribution. Then, sampling $\smash{(\mathbf{B}, \Aux)} = \smash{(B^{1:M}, \Aux)} \sim \smash{\resDistAlt(\ccdot; \WeiAll^\varpi)}$---and subsequently replacing the $m$th generated ancestor index $\smash{B^m}$ by $\smash{A^m \coloneqq \varpi(B^m)}$---induces another resampling scheme $\resScheme$ with
  \begin{align}
    \resDist (\ancAll, \aux)
    & \coloneqq \smash{\sum_{\mathbf{b} \in [N]^M} \resDistAlt(\mathbf{b}, \aux; \WeiAll^\varpi) \delta_{\varpi(\mathbf{b})}(\ancAll)
  = \resDistAlt(\varpi^{-1}(\ancAll), \aux; \WeiAll^\varpi),}\vphantom{\prod_i}\\
    \indDist (m, \aux|n)
    & \coloneqq \frac{\resDistAlt^m(\varpi^{-1}(n); \WeiAll^\varpi)}{\sum_{l=1}^M \resDistAlt^l(\varpi^{-1}(n); \WeiAll^\varpi)} \resDistAlt^m(\aux| \varpi^{-1}(n); \WeiAll^\varpi) = \kappa(m, \aux|\varpi^{-1}(n), \WeiAll^\varpi),
  \end{align}
where, for any $\smash{\mathbf{b} \coloneqq b^{1:N} \in [N]^N}$ and any permutation $\sigma \in \permutations_N$, we write $\smash{\sigma(\mathbf{b}) \coloneqq (\sigma(b^1), \dotsc, \sigma(b^N))}$. By assumption on $\resSchemeAlt$, $\resScheme$ is then balanced (and hence properly weighted) and persistent.

\begin{figure}[htb]
\algorithmSepAbove
\begin{algorithm}[H]
\caption{resampling with weight re-ordering}\label{alg:resampling_with_weight_reordeing}
\begin{algorithmic}[1]
  \Require a permutation $\varpi \in \permutations_N$ which may depend on $(\StaAll, \WeiAll)$.
  \State Sample $\smash{(\mathbf{B}, \Aux) = (B^{1:M},\Aux) \sim \resDistAlt(\ccdot; \WeiAll^\varpi)}$.
  \State Set $\smash{\AncAll \coloneqq \varpi(\mathbf{B})}$.
  \State Set $\smash{\StaRes^m \coloneqq \Sta^{A^m}}$ and $\smash{\weiRes^m \coloneqq W^{A^m} / \sum_{l=1}^M \resDistAlt^l(\varpi^{-1}(A^m); \WeiAll^\varpi)}$, for $m \in [M]$.
\end{algorithmic}
\end{algorithm}
\algorithmSepBelow
\end{figure}

\begin{figure}[htb]
\algorithmSepAboveAlt
\begin{algorithm}[H]
\caption{conditional resampling with weight re-ordering}\label{alg:conditional_resampling_with_weight_re-ordering}
\begin{algorithmic}[1]
  \Require $a \in [N]$ with $\smash{W^a > 0}$; a permutation $\varpi \in \permutations_N$ which may depend on $(\StaAll, \WeiAll)$.
  \State Sample $\smash{(K, \mathbf{B}, \Aux) \sim \kappa(k, \aux|\varpi^{-1}(a); \WeiAll^\varpi) \delta_{\varpi^{-1}(a)}(b^k) \resDistAlt^{-k}(\mathbf{b}^{-k}|b^k, \aux; \WeiAll^\varpi)}$.
  \State Set $\AncAll \coloneqq \varpi(\mathbf{B})$.
  \State Set $\smash{\StaRes^m \coloneqq \Sta^{A^m}}$ and $\smash{\weiRes^m \coloneqq W^{A^m} /  \sum_{l=1}^M \resDistAlt^l(\varpi^{-1}(A^m); \WeiAll^\varpi)}$, for $m \in [M]$.
\end{algorithmic}
\end{algorithm}
\algorithmSepBelow
\end{figure}

In particular, note that if $\resSchemeAlt$ is weight-homogeneous (which holds for all the elementary resampling schemes discussed in Section~\ref{subsec:elementary_resampling_schemes} except for `trivial' resampling) then $\resScheme$ is again weight-homogeneous. If $\resSchemeAlt$ is the stratified or systematic resampling scheme and if $\varpi$ is a mean partition order for $\WeiAll$, i.e., $W^{\varpi(1)}, \dotsc, W^{\varpi(N_0)} \leq \frac{1}{N}\sum_{n=1}^N W^n$, and $W^{\varpi(N_0)}, \dotsc, W^{\varpi(N)} > \frac{1}{N}\sum_{n=1}^N W^n$, for some $N_0 \in [N]$,
then Algorithm~\ref{alg:resampling_with_weight_reordeing} recovers the mean-order versions of stratified/systematic resampling proposed in \citet{chopin2022resampling};  \citet{karppinen2023conditional} recently justified its conditional version by using random shifts.

\subsubsection{Mixture resampling and adaptive resampling}
\label{subsubsec:adaptive_resampling}

For some $I \in \naturals$, let $\{\langle \resDistAlt_i, \indDistAlt_i\rangle\}_{i \in [I]}$ be family of balanced and persistent resampling schemes each of which may make use of some auxiliary variables $\Aux_i$ taking values in some space $\spaceU_i$. Additionally, let $V \in [I]$ be some random indicator whose distribution $\resDist(v) = \resDist(v; \WeiAll)$ may depend on the weights $\smash{\WeiAll = W^{1:N}}$ although we do not make this dependence explicit in the notation.

We will now specify a \emph{mixture resampling} scheme $\resScheme$ which performs resampling according to $\langle \resDistAlt_v, \indDistAlt_v\rangle$ with probability $\resDist(v)$: Write $\smash{\spaceU \coloneqq [I] \times \bigtimes_{i=1}^I \spaceU_i}$ as well as $\smash{\Aux \coloneqq (V, \Aux_{1:I})}$ and define
\begin{align}
    \resDist(\ancAll, \aux) & \coloneqq \resDist(v) \resDistAlt_{v}(\ancAll, \aux_{v}) \smash{\prodSubstackAligned{i}{=}{1}{i}{\neq}{v}^I \resDistAlt_i(\aux_i),}\vphantom{\prod^I}
\end{align}
and (whenever $\smash{W^n > 0}$):\\[-4ex]
\begin{align}
   \indDist(m, \aux| n) 
   & \coloneqq 
   \smash{\frac{\smash{\resDist(v) \resDistAlt_{v}^m(n, u_v)}}{\sum_{l=1}^M \resDistAlt_{v}^l(n)} \prodSubstackAligned{i}{=}{1}{i}{\neq}{v}^{\smash{I}} \resDistAlt_i(\aux_i)
   = \resDist(v) \kappa_{v}(m, \aux_{v}|n) \prodSubstackAligned{i}{=}{1}{i}{\neq}{v}^{\smash{I}} \resDistAlt_i(\aux_i).}\vphantom{\prod_{i_{i_i}}^I}
\end{align}
Then $\resScheme$ is properly weighted and persistent but not generally balanced (unless $\resDist(i) = 1$ for some $i \in [I]$). To see this, note that 
  $\weiResFun(m, n, \aux) 
  = \smash{M W^n / \sum_{l=1}^M \resDistAlt_{v}^l(n)}$
still depends on the component $v$ of $\aux$. We could have made $\resScheme$ balanced by setting $\indDist(m, \diff \aux|n) \propto \resDist^m(n, \diff \aux)$. However, then $\smash{\weiResFun(m, n, \aux) = M W^n / \sum_{i=1}^I \resDist(i) \sum_{l=1}^M \resDistAlt_{i}^l(n) }$, i.e., the cost of evaluating post-resampling weights grows with the number of mixture components, $I$.

\begin{figure}[htb]
\algorithmSepAbove
\begin{algorithm}[H]
\caption{mixture resampling}\label{alg:adaptive_resampling}
\begin{algorithmic}[1]
      \State Sample $V \sim \resDist(v)$ and $\smash{(\AncAll, \Aux_V) \sim \resDistAlt_{V}(\ancAll, \aux_V)}$;
      \State Set $\smash{\StaRes^m \coloneqq \Sta^{A^m}}$ as well as
        $\smash{\weiRes^m \coloneqq W^{\Anc^m} / \sum_{l=1}^M \resDistAlt_V^l(A^m)}$, for $m \in [M]$.
\end{algorithmic}
\end{algorithm}
\algorithmSepBelow
\end{figure}

\begin{figure}[htb]
\algorithmSepAboveAlt
\begin{algorithm}[H]
\caption{conditional mixture resampling}\label{alg:adaptive_conditional_resampling}
\begin{algorithmic}[1]
  \Require $a \in [N]$ with $\smash{W^a > 0}$.
    \State Sample $V \sim \resDist(v)$.
    \State Sample $\smash{(K, \AncAll, \Aux_V) \sim \kappa_V(k, \aux_V |a) \delta_{a}(a^k) \resDistAlt_V^{-k}(\ancAll^{-k}|\anc^k\!, \aux_V)}$.
    \State Set $\smash{\StaRes^m \coloneqq \Sta^{A^m}}$ as well as
        $\smash{\weiRes^m \coloneqq W^{\Anc^m} / \sum_{l=1}^M \resDistAlt_V^l(A^m)}$, for $m \in [M]$.
\end{algorithmic}
\end{algorithm}
\algorithmSepBelow
\end{figure}

A special case of such a mixture resampling scheme is frequently employed within \emph{\gls{SMC}} methods (to be discussed in Section~\ref{sec:application_to_smc_methods}). Assume that $M = N$, $I = 2$, and that $\langle \resDistAlt_2, \indDistAlt_2 \rangle$ is the `trivial' resampling scheme from Section~\ref{subsubsec:trivial_resampling}. A common choice, first proposed in \citet{liu1995blind}, is then to decide `whether or not to resample' based on some $\calF$-measurable function such as the \emph{\gls{ESS}} of the weights \citep{kong1994sequential}:
\begin{align}
    \ess(\WeiAll) \coloneqq \smash{\biggl[\sum_{n=1}^N (W^n)^2\biggr]^{-1},} \vphantom{\biggl[\biggr]}
\end{align}
which corresponds to the degenerate distribution
\begin{align}
  \resDist(v) = \resDist(v; \WeiAll) \coloneqq \mathbf{1}\{\ess(\WeiAll) \leq \eta N\} \delta_1(v) + \mathbf{1}\{\ess(\WeiAll) > \eta N\} \delta_2(v),
\end{align}
for some user-defined threshold $\eta \in [0, 1]$. In the remainder of this work, we refer to this special case of mixture resampling as \emph{\gls{ESS}-adaptive resampling.}

\subsubsection{Partitioned resampling, partial resampling and optimal finite-state resampling}
\label{subsubsec:optimal_finite-state_resampling}

Let $I \in \naturals$ and for $i \in [I]$, let $\smash{L_i \coloneqq \{l_i^1, \dotsc, l_i^{N_i}\} \subseteq [N]}$, be such that $\smash{\{L_1, \dotsc, L_I\}}$ is a partition of $[N]$. 
Then we can `partition' the weighted sample $(\StaAll, \WeiAll)$ into $I$ disjoint blocks of weighted samples of the form $\smash{(\mathbf{X}_i, \mathbf{W}_i) \coloneqq (X_i^{1:N_i}, W_i^{1:{N_i}})}$, where for all $n \in [N_i]$:
\begin{align}
  X_i^n \coloneqq X^{l_i^n}, \qquad \text{and} \qquad W_i^n \coloneqq \smash{\frac{W^{l_i^n}}{\sum_{m=1}^{N_i} W^{l_i^m}}.}\vphantom{\prod_i^I}
\end{align}
For $i \in [I]$, let $\langle \resDistAlt_i, \indDistAlt_i \rangle$ be some balanced and persistent resampling scheme which generates $M_i$ ancestor indices $\smash{\AncAll_i = \Anc_i^{1:M_i} \in [N_i]^{M_i}}$ for the weighted sample $(\mathbf{X}_i, \mathbf{W}_i)$ and may make use of some auxiliary variables $\Aux_i$ taking values in some space $\spaceU_i$.

We will now specify a \emph{partitioned resampling} scheme $\resScheme$ which generates ancestor indices $\smash{\AncAll \in \bigtimes_{i=1}^I L_i^{M_i}}$ from $\AncAll_1, \dotsc, \AncAll_I$ by setting
\begin{align}
  \smash{\AncAll \coloneqq h(\AncAll_1, \dotsc, \AncAll_I) \coloneqq (l_1^{A_1^1}, \dotsc, l_1^{A_1^{M_1}}, l_2^{A_2^1}, \dotsc, l_2^{A_2^{M_2}}, \dotsc, l_I^{A_I^1}, \dotsc, l_I^{A_I^{M_I}}).}
\end{align}
More precisely, we set $\smash{M \coloneqq \sum_{i=1}^I M_i}$, $\smash{\spaceU \coloneqq \bigtimes_{i=1}^I \spaceU_i}$, $\smash{\Aux \coloneqq  \Aux_{1:I}}$ and
\begin{align}
    \smash{\resDist(\ancAll, \aux)}
    & \coloneqq
    \smash{\sum_{(\ancAll_1, \dotsc, \ancAll_I) \in \bigtimes_{i=1}^I [N_i]^{M_i}} \delta_{h(\ancAll_1, \dotsc, \ancAll_I)}(\ancAll)
    \prod_{i=1}^I
    \resDistAlt_i(\ancAll_i, \aux_i),} \vphantom{\prod_i^i}\\
   \smash{\indDist(m, \aux| n)}
    & \coloneqq
   \smash{\frac{\resDist^m(n, \aux)}{\sum_{m=1}^M \resDist^m(n)}
   = \kappa_{i(n)}\Bigl(m - \smashoperator{\sum_{j=1}^{i(n)-1}} M_j, \aux_{i(n)} \,\Big|\,q(n)\Bigr) \prodSubstackAligned{j}{=}{1}{j}{\neq}{i(n)}^I \varrho_j(\aux_j),} \vphantom{\prod_{I_I}^i}\\
  \weiResFun(m, n, \aux) 
  & = \frac{\Wei^n \indDist(m, \aux| n)}{M^{-1}\resDist^m(n, \aux)}
  = \frac{M W^n}{\sum_{l=1}^{M_{i(n)}} \varrho_{i(n)}^l(q(n))},
\end{align}
where $i(n) \in [I]$ and $q(n) \in \smash{[N_{i(n)}]}$ are the unique solutions to $\smash{n \in L_{i(n)}}$ and $\smash{n = l_{i(n)}^{q(n)}}$, respectively. Then $\resScheme$ is again balanced (and thus properly weighted) and persistent.

\begin{figure}[htb]
\algorithmSepAbove
\begin{algorithm}[H]
\caption{partitioned resampling}\label{alg:partitioned_resampling}
\begin{algorithmic}[1]
\State Sample $\smash{(\AncAll_i, \Aux_i) \sim \resDistAlt_i(\ancAll_i, \aux_i; \WeiAll_i)}$, for $i \in [I]$.
      \State Set $\smash{\AncAll \coloneqq h(\AncAll_1, \dotsc, \AncAll_I)}$.
      \State Set $\smash{\StaRes^m \coloneqq \Sta^{A^m}}$ as well as
        $\smash{\weiRes^m \coloneqq W^{\Anc^m} / \sum_{l=1}^{M_{i(A^m)}} \varrho_{i(A^m)}^l(q(A^m))}$, for $m \in [M]$.
\end{algorithmic}
\end{algorithm}
\algorithmSepBelow
\end{figure}

\begin{figure}[htb]
\algorithmSepAboveAlt
\begin{algorithm}[H]
\caption{conditional partitioned resampling}\label{alg:conditional_partitioned_resampling}
\begin{algorithmic}[1]
  \Require $a \in [N]$ with $\smash{W^a > 0}$.
 \State Sample $$\smash{(L, \AncAll_{i(a)}, \Aux_{i(a)}) \sim \kappa_{i(a)}(l, \aux_{i(a)} | q(a); \WeiAll_{i(a)}) \delta_{q(a)}(a_{i(a)}^l) \resDistAlt_{i(a)}^{-l}(\ancAll_{i(a)}^{-l}|a_{i(a)}^l, \aux_{i(a)}; \WeiAll_{i(a)})}$$ as well as $\smash{(\AncAll_j, \Aux_j) \sim \resDistAlt_j(\ancAll_j, \aux_j; \WeiAll_j)}$, for all $j \in [I] \setminus \{i(a)\}$.
  \State Set $\smash{\AncAll \coloneqq h(\AncAll_1, \dotsc, \AncAll_I)}$ and $\smash{K \coloneqq L + \sum_{j=1}^{i(a)-1} M_j}$.
  \State Set $\smash{\StaRes^m \coloneqq \Sta^{A^m}}$ as well as
    $\smash{\weiRes^m \coloneqq W^{\Anc^m} / \sum_{l=1}^{M_{i(A^m)}} \varrho_{i(A^m)}^l(q(A^m))}$, for $m \in [M]$.
\end{algorithmic}
\end{algorithm}
\algorithmSepBelow
\end{figure}

We end this section with two special cases of partitioned resampling that have been proposed in the literature. Let $I = 2$ and let $\langle \resDistAlt_1, \indDistAlt_1 \rangle$ be the trivial resampling scheme from Section~\ref{subsubsec:trivial_resampling} (with $M_1 = N_1$) and let $\langle \resDistAlt_2, \indDistAlt_2 \rangle$ be some other weight-homogeneous (for simplicity) resampling scheme.

Then $\resScheme$ reduces to the \emph{partial resampling} scheme from \citet{martino2016weighting} (a related method was proposed in \citealt{bolic2004resampling}) which resamples only the particles with indices in $L_2$ but keeps those with indices in $L_1$ unchanged. Weight-homogeneity of $\smash{\langle \resDistAlt_2, \indDistAlt_2 \rangle}$ then implies $\smash{\sum_{m=1}^{M_2} \varrho_{2}^m(n) = M_2 W_2^n}$, for any $n \in [N_2]$, so that
\begin{align}
  \weiResFun(m, n, \aux)
  & = \frac{MW^n}{\sum_{l=1}^{M_{i(n)}} \varrho_{i(n)}^l(q(n))}
  = 
  \begin{dcases}
    \smash{MW^n}, & \text{if $i(n) = 1$,}\\
    \frac{\smash{M W^n}}{M_2 W_2^{q(n)}} = \frac{M}{M_2} \sum_{j=1}^{\smash{N_2}} W^{l_2^j}, & \text{if $i(n) = 2$.}
  \end{dcases} \label{eq:partial_resampling_post_resampling_weights}
\end{align}

A sophisticated version of such a partial resampling scheme known as \emph{optimal finite-state resampling} was proposed in \citet{fearnhead1998sequential, fearnhead2003online}; and its conditional version in \citet{whiteley2010efficient}. It uses \citet[][Algorithm~5.2]{fearnhead1998sequential} to solve
\begin{align}
 \sum_{n=1}^{N} \min\{1, C W^n\} = M, \label{eq:implicit_definition_of_c}
\end{align}
for $C > 0$ and then sets $\smash{L_1} \coloneqq \smash{\{n \in [N] \mid C W^n \geq 1\}}$ and $L_2 \coloneqq \smash{\{n \in [N] \mid C W^n < 1\}}$. By \eqref{eq:implicit_definition_of_c}, we then have $\smash{\frac{1}{M_2} \sum_{n=1}^{N_2} W^{l_2^n} = 1 / C}$ so that \eqref{eq:partial_resampling_post_resampling_weights} takes the following simple form:
 $\smash{\weiResFun(m, n, \aux)}
  = \smash{M W^n / \min\{1, C W^n\}}$.
If $M \leq N$, and if $\langle \resDistAlt_2, \indDistAlt_2\rangle$ is stratified or systematic resampling, this scheme ensures that the family of ancestor indices $\AncAll$ contains no duplicates. This minimises the variance induced by the resampling operation when $\spaceX$ is finite which is why optimal finite-state resampling is at the heart of the \emph{discrete particle filters} from \citet{fearnhead1998sequential, fearnhead2003online} and related \gls{SMC} and \gls{CSMC} algorithms for change-point models \citep{fearnhead2007online, whiteley2009particle}.

\subsubsection{Modified-weight and chopthin resampling}
\label{subsubsec:chopthin_resampling}

Consider a weight-homogeneous resampling scheme which generates $M$ ancestor indices using weights $\Wei^{1:N}$. Then by the conventional `unbiasedness' condition \eqref{eq:conventional_unbiasedness_condition}, the expected number of descendants of the $n$th particle is $M \smash{W^n}$. More generally, we may wish to enforce that the expected number of descendants of the $n$th particle is $\smash{H^n} \neq M\smash{W^n}$, for some $\smash{H^1, \dotsc, H^N} \geq 0$ such that  
\begin{align}
\forall\, n \in [N]:  H^n > 0 \text{ if and only if } W^n > 0; \qquad \text{and} \qquad  \sum_{n=1}^N H^n = M. \label{eq:modified_weight_resampling_assumption}
\end{align}
This can be achieved as follows. Let $\resSchemeAlt$ be some weight-homogeneous resampling scheme which generates $M$ ancestors based on the weights $\smash{V^{1:N}}$, where $\smash{V^n = H^n / M}$. Then, taking 
\begin{align}
  \resDist(\ancAll, \aux) 
   \coloneqq \smash{\resDistAlt(\ancAll, \aux; V^{1:N})}, \qquad
  \indDist(m, \aux|n)
   \coloneqq \smash{\indDistAlt(m, \aux|n; V^{1:N})},\label{eq:modified_weight_resampling}
\end{align}
i.e., $\weiResFun(m,n,\aux) = \smash{M W^n / H^n}$, gives a generally no longer weight-homogeneous but still balanced and persistent resampling scheme $\resScheme$ which satisfies \eqref{eq:modified_weight_resampling_assumption}. Resampling according to alternative weights has a long history (e.g., \citet{pitt1999filtering, liu2001monte, chen2010particle}).

\begin{figure}[htb]
\algorithmSepAbove
\begin{algorithm}[H]
\caption{modified-weight resampling}\label{alg:modified_weight_resampling}
\begin{algorithmic}[1]
  \Require $H^{1:N}$ satisfying \eqref{eq:modified_weight_resampling_assumption}.
   \State Sample $(\AncAll, \Aux)$ via some weight-homogeneous resampling scheme (e.g., via Algorithms~\ref{alg:multinomial_resampling}, \ref{alg:stratified_resampling}, \ref{alg:systematic_resampling} or \ref{alg:residual_resampling}) but using weights $\smash{V^n \coloneqq H^n / M}$ instead of $\smash{W^n}$, for $n \in [N]$.
   \State Set $\smash{\StaRes^m \coloneqq \Sta^{A^m}}$ and $\smash{\weiRes^m \coloneqq W^{A^m}/H^{A^m}}$, for $m \in [M]$. \label{alg:modified_weight_resampling:2}
\end{algorithmic}
\end{algorithm}
\algorithmSepBelow
\end{figure}

\begin{figure}[htb]
\algorithmSepAboveAlt
\begin{algorithm}[H]
\caption{conditional modified-weight resampling}\label{alg:conditional_modified_weight_resampling}
\begin{algorithmic}[1]
  \Require $a \in [N]$ with $\smash{W^a > 0}$; $H^{1:N}$ satisfying \eqref{eq:modified_weight_resampling_assumption}.
   \State Sample $(K, \AncAll, \Aux)$ via the conditional version of some weight-homogeneous resampling scheme (e.g., via Algorithms~\ref{alg:conditional_multinomial_resampling},  \ref{alg:conditional_stratified_resampling}, \ref{alg:conditional_systematic_resampling} or \ref{alg:conditional_residual_resampling}) but using weights $\smash{V^n \coloneqq H^n / M}$ instead of $\smash{W^n}$, for $n \in [N]$. \label{alg:conditional_modified_weight_resampling:1}
   \State Set $\smash{\StaRes^m \coloneqq \Sta^{A^m}}$ and $\smash{\weiRes^m \coloneqq W^{A^m}/H^{A^m}}$, for $m \in [M]$. \label{alg:conditional_modified_weight_resampling:2}
\end{algorithmic}
\end{algorithm}
\algorithmSepBelow
\end{figure}

As an example of this idea, \citet{gandy2015chopthinv3} sought a resampling scheme such that:
\begin{enumerate}
 \item\label{enum:chopthin:1} particles whose weights are below some threshold $\alpha > 0$ have zero or one descendants (`thin'); 
 \item\label{enum:chopthin:2} particles whose weights are equal to or above $\alpha$ have one or more offspring (`chop');
 \item\label{enum:chopthin:3} the variability of the post-resampling `weights' can be controlled, i.e., for some $\tilde{\beta} \geq 1$ chosen by the user: $\smash{\max_{(k,l) \in [M]^2}\weiRes^k / \weiRes^l \leq \tilde{\beta}}$.
\end{enumerate}
These goals could be ensured by Algorithm~\ref{alg:modified_weight_resampling} or \ref{alg:conditional_modified_weight_resampling} with systematic resampling as the weight-homogeneous resampling scheme  $\resSchemeAlt$ and $\smash{H^n} \coloneqq \smash{h_{\alpha^{\mathrlap{*}},\beta}(W^n)}$ defined through the continuous mapping 
\begin{align}
  [0,1] \ni W \mapsto h_{\alpha, \beta}(W) \coloneqq
  \begin{cases}
    W / \alpha, & \text{if $W < \alpha$,}\\
    1, & \text{if $\alpha \leq W < \alpha \beta / 2$,}\\
    2 W / (\alpha \beta), & \text{if $\alpha \beta / 2 \leq W$.}
  \end{cases}
\end{align}
Above, given $\beta \geq 4$, $\alpha^* \in (0,1]$---found via \citet[][Algorithm~2]{gandy2015chopthinv3}---solves
\begin{align}
  \forall\, n \in [N]:  h_{\alpha^{\mathrlap{*}},\beta}(W^n) > 0 \text{ if and only if } W^n > 0; \qquad \text{and} \qquad  \sum_{n=1}^N h_{\alpha^{\mathrlap{*}}, \beta}(W^n) = M. \label{eq:chopthin_condition_on_alpha}
\end{align}
Thus, $\smash{\weiRes^m = W^{A^m} / H^{A^m}}$, i.e., $\weiRes^m = \alpha^*$, if $\smash{W^{A^m} < \alpha^*}$, and $\smash{\weiRes^m = \min\{W^{A^m}, \alpha^* \beta / 2\}}$, if $\smash{W^{A^m} \geq \alpha^*}$. 

However, \citet{gandy2015chopthinv3} instead set $\smash{\weiRes^m \coloneqq W^{A^m} / \sum_{l=1}^M \ind\{A^l = A^m\}}$, if $\smash{W^{A^m} \geq \alpha^*}$. We now justify this within our framework using an enlarged set of auxiliary variables and an imbalanced index distribution.
Write $\aux \coloneqq (\aux_0, \mathbf{b})\in \smash{\spaceU_0 \times [N]^M \eqqcolon \spaceU}$ with $\aux_0 \in \spaceU_0$ and $\mathbf{b} \coloneqq \smash{b^{1:M} \in [N]^M}$ being, respectively, the auxiliary variables and ancestor indices generated by $\smash{\resDistAlt(\ccdot; V^{1:N})}$ and set: 
\begin{align}
  \resDist(\ancAll, \aux) 
  & \coloneqq \smash{\resDistAlt(\aux; V^{1:N})} \delta_{\mathbf{b}}(\ancAll), \label{eq:chopthin_resampling:1}\\
  \indDist(m, \aux|n)
  & \coloneqq 
  \begin{cases}
    \frac{\resDist^m(n, u)}{\sum_{l=1}^M \resDist^l(n)} = \smash{\indDistAlt(m, \aux_0|n; V^{1:N}) \delta_{n}(b^m) \resDistAlt^{-m}(\mathbf{b}^{-m}|b^m, \aux_0; V^{1:N})}, & \text{if $\smash{W^n < \alpha^*}$,}\\
    \smash{\resDistAlt(\aux; V^{1:N}) \dUnif_{\{k \in [M]\mid b^k = n\}}(m)}, & \text{if $\smash{W^n \geq \alpha^*}$,}
  \end{cases}
  \label{eq:chopthin_resampling:2}\\
  \weiResFun(m,n,\aux) 
  & = \frac{\Wei^n \indDist(m, \aux|n)}{M^{-1}\resDist^m(n,u)} = 
  \begin{cases}
    \smash{M \alpha^* \ind\{b^m = n\}}, & \text{if $\smash{W^n < \alpha^*}$,}\\
    \smash{M W^n \ind\{b^m = n\}} / \textstyle \smash{\sum_{l=1}^M \ind\{b^l = n\}}, & \text{if $\smash{W^n \geq \alpha^*}$.}
  \end{cases}\label{eq:chopthin_resampling:3}
\end{align}
For intuition, note that Particle~$n$ is guaranteed to have at least one descendant under $\resDist$ whenever $\smash{W^n \geq \alpha}$; thus, we have greater flexibility in choosing $\indDist(m,u|n)$ whenever $\smash{W^n \geq \alpha}$. 

The resampling scheme $\resScheme$, summarised in Algorithm~\ref{alg:chopthin_resampling}, is then (the version of) the \emph{chopthin} resampling scheme from \citet{gandy2015chopthinv3} which was demonstrated by the authors to outperform systematic resampling in some applications. Having cast chopthin resampling within our framework immediately shows that it is properly weighted and hence unbiased (and also persistent); it also allows us to derive the corresponding conditional version, presented in Algorithm~\ref{alg:conditional_chopthin_resampling}, which is novel.

A later version of chopthin resampling \citep{gandy2016chopthin} differs from the one of \citet{gandy2015chopthinv3} in that it additionally (a) uses weight reordering with a permutation $\varpi$ such that $W^{\varpi(1)}, \dotsc, W^{\varpi(N_0)} < \alpha^*$, and $W^{\varpi(N_0)}, \dotsc, W^{\varpi(N)} \geq \alpha^*$, for some $N_0 \in [N]$, and (b) ensures that the post-resampling weights are again normalised, i.e., $\sum_{m =1}^M \weiRes^m = 1$. The latter is achieved by adjusting the post-resampling weights as $\smash{\weiRes^m = (W^{A^m} + \xi^{A^m})/ \sum_{l=1}^M \ind\{A^l = A^m\}}$, if $\smash{W^{A^m} \geq \alpha^*}$, where $\xi^n$ is a certain random variable satisfying $\E[\xi^n|\calF] = 0$ and $\xi^n - W^n \geq 0$, almost surely. Whether (b) can be justified within our framework (e.g., in order to derive a conditional version) remains an open question.

\begin{figure}[htb]
\algorithmSepAbove
\begin{algorithm}[H]
\caption{chopthin resampling}\label{alg:chopthin_resampling}
\begin{algorithmic}[1]
   \State Find $\alpha^*$ satisfying \eqref{eq:chopthin_condition_on_alpha} via \citet[][Algorithm~2]{gandy2015chopthinv3}; set $H^n \coloneqq h_{\alpha^{\mathrlap{*}},\beta}(W^n)$, for $n \in [N]$.
   \State Sample $(\AncAll, \Aux)$ via systematic resampling (Algorithm~\ref{alg:systematic_resampling}) but using weights $V^n \coloneqq H^n / M$ instead of $W^n$, for $n \in [N]$.
   \State Set $\smash{\StaRes^m \coloneqq \Sta^{A^m}}$ and $\smash{\weiRes^m \coloneqq \ind\{W^{A^{\mathrlap{m}}} < \alpha^*\} \alpha^*  +  \ind\{W^{A^m} \geq \alpha^*\} W^{A^m}/\sum_{l=1}^M \ind\{A^l = A^m\}}$, for $m \in [M]$. \label{alg:chopthin_resampling:2}
\end{algorithmic}
\end{algorithm}
\algorithmSepBelow
\end{figure}

\begin{figure}[htb]
\algorithmSepAboveSmall
\begin{algorithm}[H]
\caption{conditional chopthin resampling}\label{alg:conditional_chopthin_resampling}
\begin{algorithmic}[1]
  \Require $a \in [N]$ with $\smash{W^a > 0}$.
   \State Find $\alpha^*$ satisfying \eqref{eq:chopthin_condition_on_alpha} via \citet[][Algorithm~2]{gandy2015chopthinv3}; set $H^n \coloneqq h_{\alpha^{\mathrlap{*}},\beta}(W^n)$, for $n \in [N]$.

   \If{$\smash{W^a < \alpha^*}$}
    \State Sample $(K, \AncAll, \Aux)$ via conditional systematic resampling (Algorithm~\ref{alg:conditional_systematic_resampling}) but using weights $V^n \coloneqq H^n / M$ instead of $W^n$, for $n \in [N]$.
   \Else
    \State Sample $(\AncAll, \Aux)$ via systematic resampling (Algorithm~\ref{alg:systematic_resampling}) but using weights $V^n \coloneqq H^n / M$ instead of $W^n$, for $n \in [N]$.
    \State Sample $K \sim \dUnif_{\{l \in [M] \mid A^l = a\}}$.
   \EndIf
   \label{alg:conditional_chopthin_resampling:1}
   \State Set $\smash{\StaRes^m \coloneqq \Sta^{A^m}}$ and $\smash{\weiRes^m \coloneqq \ind\{W^{A^{\mathrlap{m}}} < \alpha^*\} \alpha^*  +  \ind\{W^{A^m} \geq \alpha^*\} W^{A^m}/\sum_{l=1}^M \ind\{A^l = A^m\}}$, for $m \in [M]$. \label{alg:conditional_chopthin_resampling:2}
\end{algorithmic}
\end{algorithm}
\algorithmSepBelowSmall
\end{figure}

\section{Application to sequential Monte Carlo methods}
\label{sec:application_to_smc_methods}
\glsreset{SMC}
\glsreset{CSMC}

In this section, we apply our resampling framework to \emph{\gls{SMC}} \citep{gordon1993novel, delmoral2004feynman} and \emph{\gls{CSMC}} \citep{andrieu2010particle} methods \citep[see][for further references]{chopin2020introduction}. Specifically, our framework gives simple proofs that \gls{SMC} and \gls{CSMC} algorithms can be implemented in a valid manner even if the resampling scheme 
    (a) violates the (`unbiasedness',  exchangeability or `marginal unbiasedness') conditions commonly imposed in the literature, e.g., in \citet{andrieu2010particle, chopin2015particle, karppinen2023conditional};
    (b) has a complicated dependence structure, e.g., as in systematic or chopthin resampling; or
    (c) is adaptive.

\subsection{Sequential Monte Carlo}
\label{subsec:smc}

Let $M_t(\ccdot|\sta_{t-1})$ be conditional distributions on $\spaceX$ (`mutation kernels') and $\smash{G_t\colon \spaceX^2 \to (0, \infty)}$ (`potential functions'). Our goal is to estimate expectations w.r.t.\ the distribution $\pi_T(\diff \sta_{1:T})\coloneqq \gamma_T(\diff \sta_{1:T}) / \normConst_T$, where
    $\smash{\gamma_T(\diff \sta_{1:T}) 
    \coloneqq \prod_{t=1}^T G_t(\sta_{t-1:t}) M_t(\diff \sta_t|\sta_{t-1})}$,
    with typically unknown normalising constant $\smash{\normConst_T}
    \coloneqq \smash{\int_{\spaceX^T} \gamma_T(\diff \sta_{1:T})}$; we use the convention that any quantity with subscript $t < 1$ is to be ignored, i.e., $M_1(\diff \sta_1|\sta_0) = M_1(\diff \sta_1)$ and $G_1(\sta_{0:1}) = G_1(\sta_1)$.

The \gls{SMC} algorithm is outlined in Algorithm~\ref{alg:smc}. It uses $N_t \in \naturals$ particles at time $t$. For simplicity, we assume that $N_t$ is deterministic but random numbers of particles could easily be accommodated by further extending the state space to explicitly include $N_t$. Additionally, we let $\indDist_1$ be some distribution on $[N_1]$ (usually, $\indDist_1 \coloneqq \dUnif_{[N_1]}$). For any $t > 1$, we let $\resSchemeTime{t}$ be a properly weighted and persistent resampling scheme which draws $\smash{N_t}$ ancestors based on $N_{t-1}$ weights $\WeiAll_{t-1} = \smash{W_{t-1}^{1:N_{t-1}}}$ that are defined below and which specifies post-resampling weights via
\begin{align}
  \weiResFun_t(m, n, \aux; \WeiAll_{t-1}) \coloneqq
 \dfrac{W_{t-1}^{n}\indDist_t(m, \aux|n; \WeiAll_{t-1})}{N_t^{-1}\resDist_{t-1}^m(n, \aux; \WeiAll_{t-1})}. \label{eq:smc_post_resampling_weight_function}
\end{align}

\begin{figure}[htb]
\algorithmSepAboveSmall
\begin{algorithm}[H]
\caption{\gls{SMC}}\label{alg:smc}
\begin{algorithmic}[1]
  \For{$t = 1,\dotsc, T$}
  \If{$t = 1$}
    \State Set $\weiRes_{t-1}^n \coloneqq \indDist_1(n)$, for $n \in [N_1]$.
  \Else
    \State Sample $\smash{(\AncAll_{t-1}, \Aux_{t-1}) = (A_{t-1}^{1:N_t}, \Aux_{t-1}) \sim \resDist_{t-1}(\ccdot; \WeiAll_{t-1})}$.
    \State Set $\smash{\weiRes_{t-1}^n \coloneqq \weiResFun_t(n, \Anc_{t-1}^n, \Aux_{t-1}; \WeiAll_{t-1}) / N_t}$ via \eqref{eq:smc_post_resampling_weight_function}, and $\smash{\StaRes_{t-1}^n \coloneqq \Sta_{t-1}^{A_{t-1}^n}}$, for $n \in [N_t]$.
  \EndIf
  \State Sample $\smash{\Sta_t^n \sim M_t(\ccdot|\StaRes_{t-1}^n)}$ for $n \in [N_t]$.
  \State Set $\smash{w_t^n \coloneqq \weiRes_{t-1}^n G_t(\StaRes_{t-1}^n, \Sta_t^n)}$, and $\smash{W_t^n \coloneqq w_t^n / \sum_{m=1}^{N_t} w_t^m}$, for $n \in [N_t]$.
\EndFor
\end{algorithmic}
\end{algorithm}
\algorithmSepBelow
\end{figure}

At the end of Algorithm~\ref{alg:smc}, expectations $\pi_T(\testFun_T) $, integrals $\gamma_T(\testFun_T) = \normConst_T \pi_T(\testFun)$, for some suitably integrable function $\testFun_T \colon \spaceX^T \to \reals$, and the normalising constant $\normConst_T$ can be approximated as
\begin{align}
  \hat{\pi}_T(\testFun_T) \coloneqq  \sum_{n=1}^{N_T} W_T^n \testFun_T(\Sta_{1:T}^{(n)}), \qquad \hat{\gamma}_T(\testFun_T) \coloneqq \widehat{\normConst}_T \hat{\pi}_T(\testFun_T), \qquad  \widehat{\normConst}_T & \coloneqq \prod_{t=1}^T \sum_{n=1}^{N_t} w_t^n,
\end{align}
where $\smash{\Sta_{1:t}^{(n)}}$ is recursively defined as $\smash{\Sta_{1}^{(n)} \coloneqq \Sta_1^n}$ and $\smash{\Sta_{1:t}^{(n)} \coloneqq (\Sta_{1:t-1}^{(A_{t-1}^n)}, \Sta_t^n)}$, for $t > 1$.

The following result (proved in Appendix~\ref{app:subsec:proof_of_smc_validity}) generalises the unbiasedness of $\alpha$\gls{SMC} from \citet[][Theorem~1(1)]{whiteley2016role} which uses a framework for resampling that encompasses a number of the (adaptive) resampling schemes from Section~\ref{sec:examples} but not resampling schemes with a complex dependence structure like (adaptive) systematic resampling or chopthin resampling (Appendix~\ref{app:subsec:alpha_smc_resampling} shows that $\alpha$\gls{SMC} resampling is a strict special case of our framework).
\begin{proposition}\label{prop:smc_validity}
  Assume that the resampling schemes $\resSchemeTime{t}$ are properly weighted and persistent for any $t \in [T] \setminus \{1\}$ and let $\testFun_T\colon \spaceX^T \to \reals$ be some $\smash{\pi_T}$-integrable test function. Then $\E[\hat{\gamma}_T(\testFun_T)] = \gamma_T(\testFun_T)$ and, in particular for $\testFun_T \equiv 1$, $\E[\widehat{\normConst}_T] = \normConst_T$.
\end{proposition}

%

The following simple example illustrates that proper weighting/unbiasedness of the resampling scheme is not sufficient to ensure convergence of estimators.

\begin{example}[`bad' unbiased resampling scheme]
  \label{ex:counterexample}
  For simplicity, let $N_1 = \dotsc = N_T = N$. Consider a resampling scheme $\resSchemeTime{t}$ with $\smash{\resDist_{t-1}(\ancAll_{t-1}) \coloneqq W_{t-1}^{a_{t-1}^1} \prod_{m=2}^N \delta_{a_{t-1}^1}(a_{t-1}^m)}$ (with $\spaceU = \emptyset$). This resampling scheme corresponds to selecting a single ancestor in proportion to its weight and assigning it to all particles in the current generation. If we take $\indDist_t(m|n) \coloneqq \dUnif_{[N]}(m)$, then this resampling scheme is also balanced and hence properly weighted as well as persistent. However, it is not the case that $\hat{\gamma}_T(\testFun_T)$ converges in probability to $\gamma_T(\testFun_T)$, as $N \uparrow \infty$, even though this estimate is unbiased.

  Take, for example, $T = 2$, $a, b > 0$,  $M_1(\diff \sta_1) = \dN(\diff \sta_1; 0, a)$, $M_2(\diff \sta_2|\sta_1) = \dN(\diff \sta_2; \sta_1, b)$, $G_1 \equiv 1$, $G_2 \equiv 1$ and $\testFun_T(\sta_{1:T}) = \sta_T$, so that $\gamma_T(\testFun_T) = 0$. Then one can attain via straightforward manipulations that $\hat{\gamma}_T(\testFun_T) \sim \dN(0, a + \tfrac{b}{N})$, for any $N \in \naturals$. Thus, while $\hat{\gamma}_T(\testFun_T)$ is unbiased, its variance is greater than or equal to $a$ for any $N \in \naturals$.
    
\end{example}

\subsection{Conditional sequential Monte Carlo}
\label{subsec:csmc}

Hereafter, assume that $N_t \geq 2$, for any $t \in [T]$. Then sampling $\smash{\Sta_{1:T}'}$ conditional on $\smash{\sta_{1:T}}$ via the \gls{CSMC} algorithm from Algorithm~\ref{alg:csmc_with_ancestor_tracing} induces a Markov kernel $P$ on $\smash{\spaceX^T}$. The following result is proved in Appendix~\ref{app:subsec:proof_of_csmc_validity}.

\begin{figure}[htb]
\algorithmSepAboveSmall
\begin{algorithm}[H]
\caption{\gls{CSMC}}\label{alg:csmc_with_ancestor_tracing}
\begin{algorithmic}[1]
  \Require $\smash{\sta_{1:T} \in \spaceX^T}$ with $\smash{\pi_T(\sta_{1:T}) > 0}$.
  \For {$t = 1,\dotsc, T$}  \label{alg:csmc_with_ancestor_tracing:1a}
    \If{$t = 1$}
      \State Sample $\smash{K_1 \sim \indDist_1}$.
      \State Set $\smash{\weiRes_{t-1}^n \coloneqq \indDist_1(n)}$, for $n \in [N_1]$.
    \Else
      \State Sample $\smash{(K_t, \AncAll_{t-1}^{-K_t}, \Aux_{t-1}) \sim \indDist_t(k_t, \aux_{t-1}|K_{t-1};\WeiAll_{t-1}) \resDist_{t-1}^{-k_t}(\ancAll_{t-1}^{-k_t}| k_{t-1}, \aux_{t-1}; \WeiAll_{t-1})}$.
      \State Set $\smash{\Anc_{t-1}^{K_t} \coloneqq K_{t-1}}$.

           \State Set $\smash{\weiRes_{t-1}^n \coloneqq \weiResFun_t(n, \Anc_{t-1}^n, \Aux_{t-1}; \WeiAll_{t-1}) / N_t}$ via \eqref{eq:smc_post_resampling_weight_function}, and $\smash{\StaRes_{t-1}^n \coloneqq \Sta_{t-1}^{A_{t-1}^n}}$, for $n \in [N_t]$.
    \EndIf
      \State Set $\Sta_t^{K_t} \coloneqq \sta_t$, and sample $\Sta_t^n \sim M_t(\ccdot|\StaRes_{t-1}^n)$, for $n \in [N_t] \setminus \{K_t\}$.
      \State Set $\smash{w_t^n \coloneqq \weiRes_{t-1}^n G_t(\StaRes_{t-1}^n, \Sta_t^n)}$, and $\smash{W_t^n \coloneqq w_t^n / \sum_{m=1}^{N_t} w_t^m}$, for $n \in [N_t]$.  \label{alg:csmc_with_ancestor_tracing:1b}
   \EndFor
\State Sample $L_T \sim \dCat(\WeiAll_T)$ 
    and set $\smash{L_t \coloneqq A_t^{L_{t+1}}}$,
    for $t = T-1,\dotsc, 1$.  \label{alg:csmc_with_ancestor_tracing:2}
    \State Return $\smash{\Sta_{1:T}' \coloneqq (\Sta_1^{L_1}, \dotsc, \Sta_T^{L_T})}$.  \label{alg:csmc_with_ancestor_tracing:4}
\end{algorithmic}
\end{algorithm}
\algorithmSepBelow
\end{figure}

\begin{proposition}\label{prop:csmc_validity}
  Assume that the resampling schemes $\resSchemeTime{t}$ are properly weighted and persistent for any $t \in [T] \setminus \{1\}$. Then $P$ is $\smash{\pi_T}$-invariant.
\end{proposition}

\begin{remark} \label{rem:validity_of_csmc_in_other_settings}
  To keep the presentation simple, we have a simple form of \gls{CSMC} algorithm (sometimes called \emph{\gls{CSMC} with ancestor tracing}); but Proposition~\ref{prop:csmc_with_backward_or_ancestor_sampling_validity} from Appendix~\ref{app:subsec:csmc_with_backward_or_ancestor_sampling} shows that invariance still holds for the \emph{backward sampling} \citep{whiteley2010particle} and \emph{ancestor sampling} \citep{lindsten2014particle} extensions (Algorithms~\ref{alg:csmc_with_backward_sampling} and \ref{alg:csmc_with_ancestor_sampling} in Appendix~\ref{app:subsec:csmc_with_backward_or_ancestor_sampling}) which can improve mixing of the induced Markov chain \citep{lee2020coupled, karjalainen2025mixing}.
\end{remark}

Proposition~\ref{prop:csmc_validity} generalises \citet[][Lemma~A.4]{huggins2019sequential} which proved the validity of \gls{CSMC} algorithms based around the $\alpha$\gls{SMC} framework and thus encompasses a number of the resampling schemes from Section~\ref{sec:examples} but not resampling schemes with a complex dependence structure like (adaptive) systematic resampling or chopthin resampling nor the backward or ancestor-sampling extensions. 

The following example illustrates that proper weighting/invariance of the (conditional) resampling scheme is not sufficient for ergodicity of the \gls{CSMC} kernel. 

\begin{example}[`bad' invariant resampling scheme]
  \label{ex:counterexample_invariance} Using the (properly weighted and persistent) resampling scheme from Example~\ref{ex:counterexample}, let $P$ be the Markov kernel on $\spaceX^T$ induced by the \gls{CSMC} algorithm.
  Then $P$ is invariant. However, $\smash{\resDist_{t-1}^{-k_t}(\ancAll_{t-1}^{-k_t}|k_{t-1})} = \smash{\prod_{m=1; m \neq k_{t-1}}^N \delta_{k_{t-1}}(a_{t-1}^m)}$ so that,  whenever $\smash{\Sta_{1:T}' \sim P(\ccdot|\sta_{1:T})}$, $\Sta_t' = x_t$ with probability $1$ for any $t < T$. In other words, $P$ cannot be ergodic.
\end{example}

\subsection{A note on (over)simplification}
\label{subsec:a_note_on_oversimplification}

In the case of multinomial resampling (and other exchangeable schemes), the \gls{CSMC} algorithm remains valid if we fix $K_t \coloneqq 1$, say, instead of sampling $K_t$ from the index distribution, for $t \in [T]$. In the case of multinomial resampling, fixing $K_t \coloneqq 1$, can be implemented straightforwardly using one of the following two strategies whose outcomes are statistically equivalent:
\begin{enumerate}[label = \Roman*., ref = \Roman*]
    \item\label{enum:naive:i} Draw $\smash{N_t}$ ancestor indices via standard multinomial resampling and then overwrite the first ancestor index with $1$.
    \item\label{enum:naive:ii} Set the first ancestor index to $1$ and then draw the remaining ancestor indices via standard multinomial resampling (with $\smash{N_t-1}$ instead of $\smash{N_t}$ descendants).
\end{enumerate}
These strategies are appealing because they allow for existing implementations of `unconditional' resampling to be used for conditional resampling. One might thus hope that they are valid more generally.

Unfortunately, this is not the case. For instance, as we illustrate in Appendix~\ref{app:sec:on_naive_implementations_of_conditional_resampling}, conditional systematic and residual-multinomial resampling implemented using the Strategies~\ref{enum:naive:i} or \ref{enum:naive:ii} can induce a bias in the sense that the \gls{CSMC} algorithm no longer leaves $\smash{\pi_T}$ invariant. Further, these na\"ive implementations can also worsen the autocorrelation of the Markov chain induced by the \gls{CSMC} algorithm.

\section{Discussion}
\glsreset{CSMC}
\glsreset{SMC}


We have presented a simple framework for guaranteeing the validity of resampling schemes within \gls{SMC} algorithms and their conditional analogues within \gls{CSMC} algorithms under minimal assumptions, covering almost all standard resampling mechanisms as well as more exotic approaches such as chopthin resampling, whose conditional version we derive for the first time. Empirical results demonstrate that neglecting careful conditional resampling can lead to substantial bias. For simplicity, we have focused on basic \gls{SMC} and \gls{CSMC} algorithms. However, our arguments straightforwardly extend to models with non-Markovian dynamics, potential functions which depend on the entire history of the state sequence, resampling schemes with a random number of particles, and to the related algorithms from \citet{lee2011auxiliary, delmoral2015alive, finke2016embedded, finke2023conditional, corenflos2025auxiliary, corenflos2024particle}, but the details have been omitted here in the interests of brevity.

\subsection*{Acknowledgments}
Large language models were used occasionally to assist with phrasing and with debugging and refining isolated sections of the accompanying code. All original ideas, mathematical derivations, and proofs are entirely the authors' own work, and the manuscript was written predominantly without AI assistance. For the purpose of open access, the authors have applied a Creative Commons Attribution (CC BY) licence to any Author Accepted Manuscript version arising from this submission. Correspondence should be addressed to Axel Finke (axel.finke@ncl.ac.uk).

\subsection*{Funding}
This work was supported by EPSRC grant no EP/K032208/1. AMJ acknowledges financial support from the Engineering and Physical Sciences Research Council (EPSRC; grants EP/R034710/1 and EP/T004134/1) and by United Kingdom Research and Innovation (UKRI) via grant EP/Y014650/1, as part of the ERC Synergy project OCEAN. Research of AL was supported by EPSRC grants EP/R034710/1 and EP/Y028783/1.

\subsection*{Data availability}
No new data were generated or analysed during this study. The GitHub repository \url{https://github.com/AxelFinke/resampling-in-conditional-smc-algorithms} provides the code for reproducing the results referenced in Section~\ref{subsec:a_note_on_oversimplification} and shown in Appendix~\ref{app:sec:on_naive_implementations_of_conditional_resampling}.


\renewcommand*{\bibfont}{\footnotesize}
\setlength{\bibsep}{3pt plus 0.3ex}
\bibliography{literature}

\appendix

\section{On na\"ive implementations of conditional resampling}
\label{app:sec:on_naive_implementations_of_conditional_resampling}

In this section, we illustrate that the two na\"ive strategies for implementing conditional resampling discussed in Section~\ref{subsec:a_note_on_oversimplification} of the main manuscript do not lead to a valid \gls{CSMC} algorithm. 
Specifically, we will illustrate that conditional systematic and conditional residual-multinomial resampling implemented using the Strategies~\ref{enum:naive:i}/\ref{enum:naive:ii} (hereafter referred to as \emph{na\"ive systematic resampling \ref{enum:naive:i}/\ref{enum:naive:ii}} and \emph{na\"ive residual-multinomial resampling \ref{enum:naive:i}/\ref{enum:naive:ii}}) can induce a bias in the sense that the \gls{CSMC} algorithm no longer leaves $\smash{\pi_T}$ invariant. Further, these na\"ive implementations can also worsen the autocorrelation of the Markov chain induced by the \gls{CSMC} algorithm.

\subsection{Binary state-space model}

We illustrate the issue on a toy binary state-space model for which it is possible to evaluate the invariant distribution and autocorrelation of the Markov chain induced by a \gls{CSMC} algorithm for small values of $N$ and $T$. The model is as follows. For any $t \in [T]$, $\varepsilon, \delta \in [0,1]$ and any $\sta_t \in \spaceX \coloneqq \{0, 1\}$, let
  $\smash{M_1}
  = \smash{(1-\varepsilon) \delta_0 + \varepsilon \delta_1}$,
  $\smash{M_{t+1}(\ccdot | \sta_t)}
  = \smash{(1-\varepsilon) \delta_{\sta_t} + \varepsilon \delta_{1 - \sta_t}}$ and
  $\smash{G_t(\sta_{t-1:t})} = \smash{G_t(\sta_t)}
  = \smash{(1-\delta) \ind\{\sta_t = 0\} + \delta \ind\{\sta_t = 1\}}$.
Additionally, we let $\smash{N_1 = \dotsc = N_T = N}$ and $\smash{\indDist_1 = \dUnif_{[N]}}$. 

\begin{figure}
  \centering
  \begin{subfigure}{\textwidth}
    \centering
    \includegraphics[scale = 0.65]{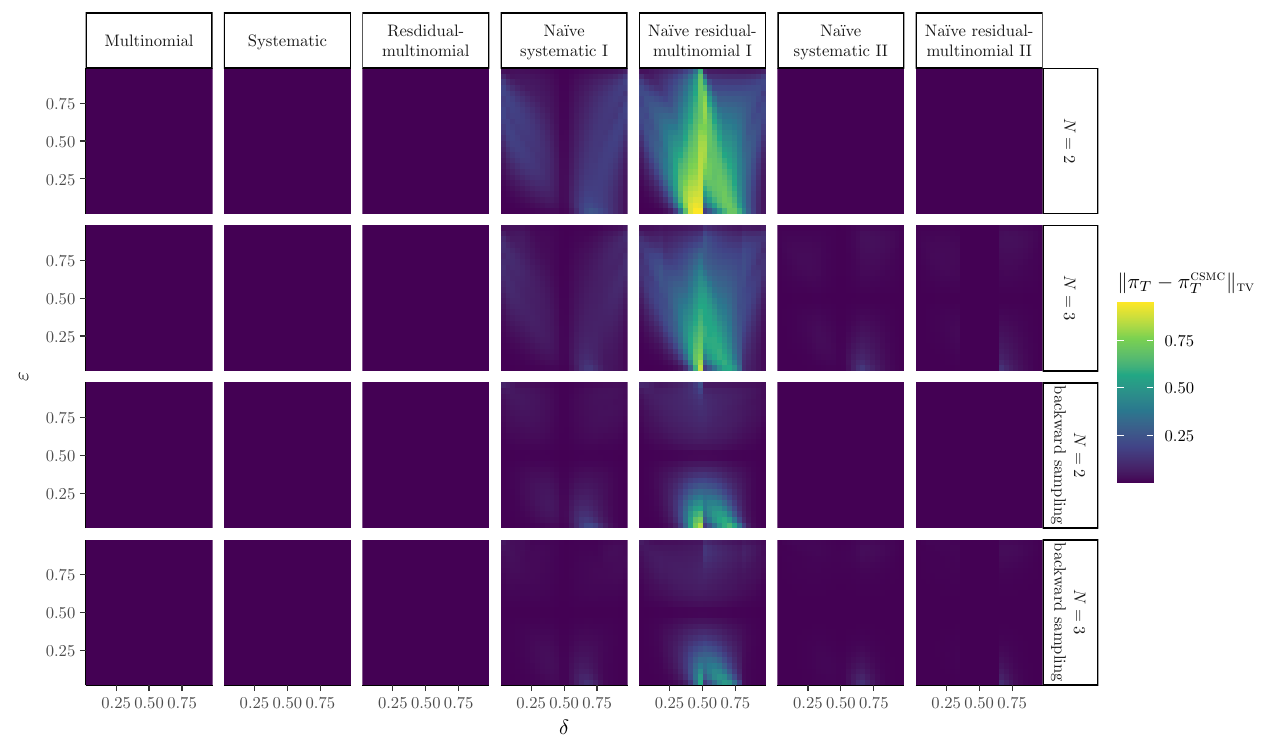}
    \caption{Bias on a linear scale.}
    \label{fig:toy_model_bias_linear_scale}
  \end{subfigure}
  \begin{subfigure}{\textwidth}
    \centering
    \includegraphics[scale = 0.65]{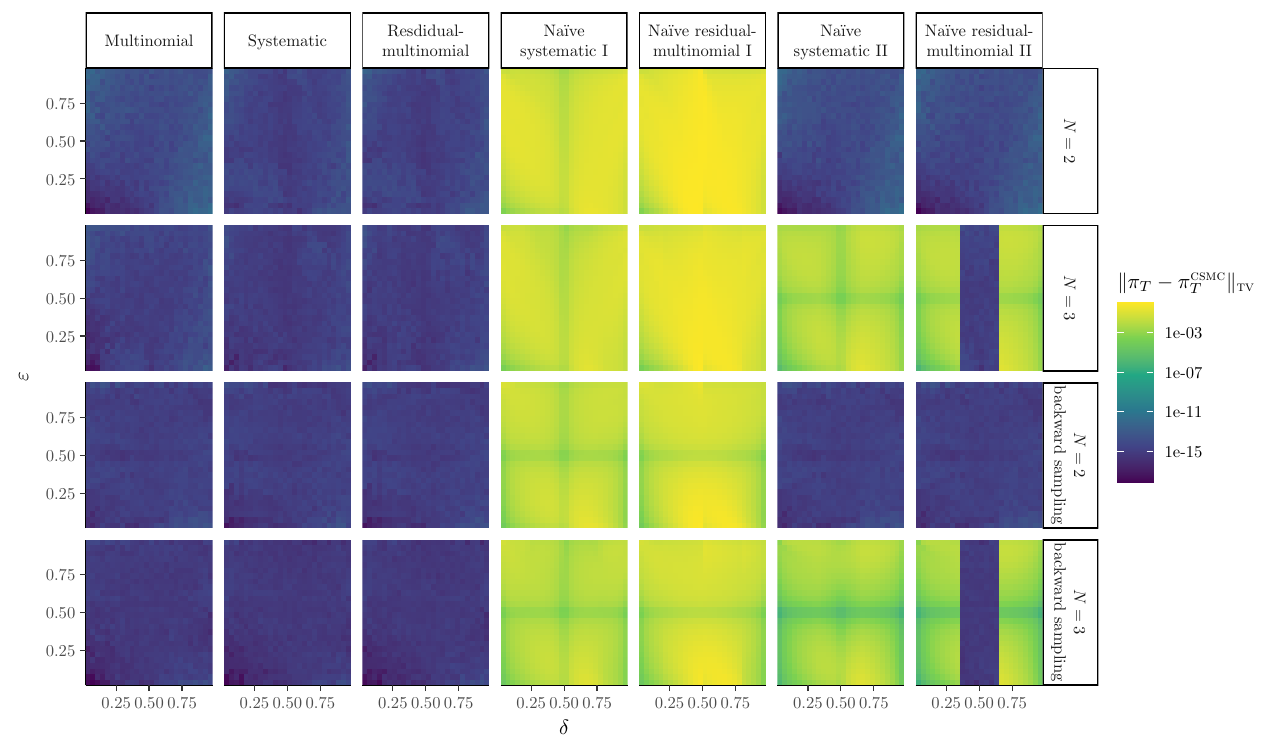}
    \caption{Bias on a logarithmic scale.}
    \label{fig:toy_model_bias_log_scale}
  \end{subfigure}
  \caption{Total variation distance between the joint smoothing distribution, $\smash{\pi_T}$, and the invariant distribution of the Markov chain induced by the \gls{CSMC} algorithm, $\pi_T^{\text{\textsc{csmc}}}$, in the binary state-space model with $T = 5$ time steps and $N_1 = \dotsc = N_T = N$ particles. This figure illustrates that a na\"ive implementation of conditional resampling can induce bias.}
   \label{fig:toy_model_bias}
\end{figure}

As illustrated in Figure~\ref{fig:toy_model_bias_linear_scale}, na\"ive conditional residual-multinomial resampling~\ref{enum:naive:i} suffers from substantial bias and increasing the number of particles or employing backward sampling reduces but does not remove this bias. Figure~\ref{fig:toy_model_bias_log_scale} demonstrates that the other na\"ive implementations also induce bias (albeit of a much smaller magnitude) with two exceptions:
\begin{enumerate}
  \item Whenever the underlying resampling scheme is unbiased and $N = 2$, the na\"ive implementation of conditional resampling using Strategy~\ref{enum:naive:ii} remains exact (in the sense that the Markov chain leaves $\smash{\pi_T}$ invariant) because it reduces to conditional multinomial resampling. We note that this holds generally, i.e., not just in this specific model.
  \item Additionally, in this specific model, it can be shown that na\"ive conditional residual-multinomial resampling~\ref{enum:naive:ii} reduces to conditional multinomial resampling (and is thus exact) whenever $\smash{\lvert \delta - \tfrac{1}{2}\rvert < 1 / (4N-6)}$.
\end{enumerate}

\begin{figure}
  \centering
    \includegraphics[scale = 0.65]{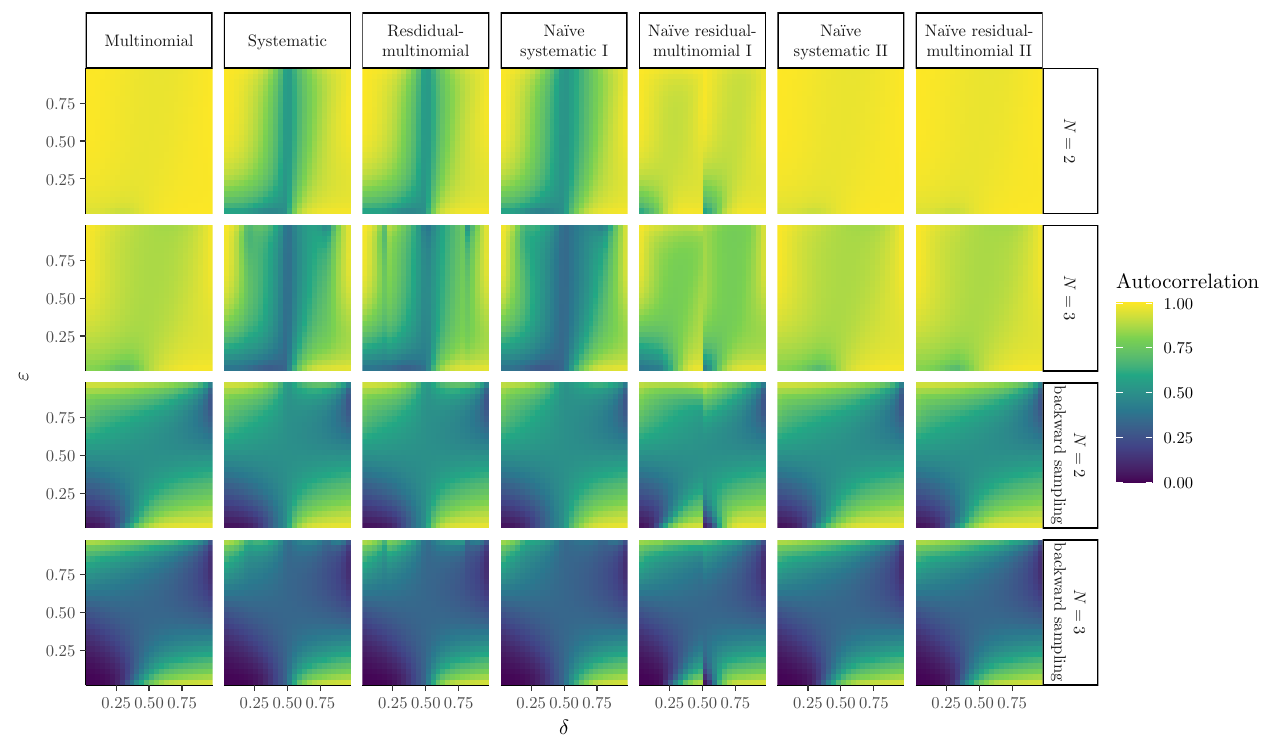}
    \caption{Lag-$1$ autocorrelation of the state at time $t = 1$ under the Markov chain induced by the \gls{CSMC} algorithm for the same setting as in Figure~\ref{fig:toy_model_bias_log_scale}. This figure illustrates that na\"ive implementations of conditional resampling can induce a substantially worse autocorrelation than their exact counterparts (except for na\"ive conditional systematic resampling~\ref{enum:naive:i}); and for nearly constant potential functions ($\delta \approx 0.5$) the difference remains even if backward sampling is used.}
    \label{fig:toy_model_autocorrelation}
\end{figure}

As illustrated in Figure~\ref{fig:toy_model_autocorrelation}, the fact that na\"ive conditional resampling \ref{enum:naive:ii} reduces to conditional multinomial resampling if $N = 2$ (and can still behave similarly to conditional multinomial resampling if $N > 2$) can also negatively impact the autocorrelation of the Markov chain induced by a \gls{CSMC} algorithm, especially when the potential functions oscillate little.

\subsection{Linear-Gaussian state-space model}

One might speculate that the extent of the bias induced by na\"ive conditional resampling \ref{enum:naive:i} was somehow due to the specific finite-state-space structure of the above example. Here, we illustrate that a similar bias is incurred in a simple univariate linear-Gaussian state-space model, i.e., if $M_1(\sta_1) = \dN(\sta_1; 0, 1)$, $M_t(\sta_t|\sta_{t-1}) = \dN(\sta_t; \sta_{t-1}, 1)$, for $t > 1$, and $G_t(\sta_{t-1:t}) = G_t(\sta_t) = \dN(y_t; \sta_t, 1)$, for given observations $y_t \in \reals$.

\begin{figure}
    \centering
    \includegraphics[scale = 0.65]{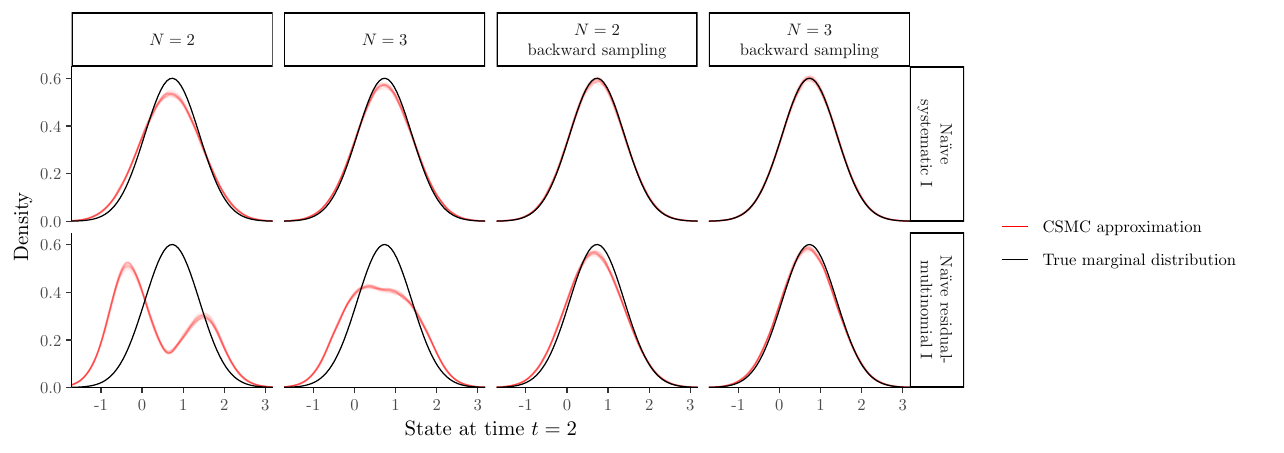}
    \caption{Approximation of the marginal smoothing distribution at time $t = 2$ in the linear-Gaussian state-space model. The red lines represent 25 independent repetitions of each algorithm using $10^6$ iterations each (after burnin).}
    \label{fig:gaussian_ssm}
\end{figure}

Figure~\ref{fig:gaussian_ssm} illustrates the bias induced in the approximation of the marginal posterior distribution at time $t=2$ by na\"ive conditional resampling \ref{enum:naive:i} in the case that $T = 4$ and $y_{1:4} = (0.75, 0.62, 1.91, 0.12)$ (the shown behaviour is typical for other observation sequences sampled from the model).

\section{Relationship with other frameworks}

\subsection{Relationship with (conditional) resampling based on randomly permuting or shifting the order of the ancestor indices}
\label{app:subsec:random_permutation}

\subsubsection{Random permutations}

The framework from \citet{andrieu2010particle} assumed, for simplicity, that the resampling schemes used within a (conditional) \gls{SMC} algorithm are exchangeable. And although many popular resampling schemes (like systematic resampling) are not exchangeable, \citet{andrieu2010particle} noted that any resampling schemes can be made exchangeable by randomly permuting the order of the ancestor indices after they have been sampled. This idea was used to derive a conditional version of residual-multinomial resampling in \citet[][Algorithm~2]{chopin2015particle} in which $K$ is always set to $1$. In this section, we show that unless an implementation of conditional resampling with $K = 1$, say, is required, the use such random permutations is unnecessary (and only adds computational cost).

Formally, let $\resSchemeAlt$ be some balanced resampling scheme which may not be exchangeable. We can then derive an exchangeable resampling scheme $\resScheme$ by setting
\begin{align}
  \resDist(\ancAll, \aux)
  = \sum_{\mathbf{b} \in [N]^M} \resDistAlt(\mathbf{b}, \aux) \sum_{\sigma \in \permutations_M} \dUnif_{\permutations_M}(\sigma) \delta_{\mathbf{b}^\sigma}(\ancAll),
\label{eq:exchangeable_resampling_distribution_through_random_permutation}
\end{align}
where for any $\mathbf{b} \coloneqq \smash{b^{1:M}}$ and any permutation $\sigma \in \permutations_M \coloneqq \smash{\{\tau\colon [M] \to [M] \mid \text{$\tau$ is bijective}\}}$, we write $\smash{\mathbf{b}^\sigma} \coloneqq \smash{(b^{\sigma(1)}, \dotsc, b^{\sigma(M)})}$; 
as well as by setting $\indDist(m, \aux|n) = \smash{\dUnif_{[M]}(m)\resDist^m(\aux|n)}$, for any $n \in [N]$; then
\begin{align}
     \weiResFun(m, n, \aux)
     & = M  W^n\frac{\indDist(m, \aux|n)}{\resDist^m(n, \aux)}\\
     & = W^n \frac{1}{\frac{1}{\# \permutations_M}\sum_{\sigma \in \permutations_M} \sum_{l=1}^M \ind\{\sigma(m) = l\} \resDistAlt^l(n)} 
     = \frac{M W^n}{\sum_{l=1}^M \resDistAlt^l(n)}.
\end{align}
Thus, application of the random permutation does not change the law of the resampled measure $\targetEstRes$.

\begin{figure}[H]
\algorithmSepAbove
\begin{algorithm}[H]
\caption{resampling with random permutation}\label{alg:resampling_with_random_permutation}
\begin{algorithmic}[1]
      \State Sample $\smash{(\mathbf{B}, \Aux) = (B^{1:M}, \Aux) \sim \resDistAlt}$.
      \State Sample $\smash{\varSigma \sim \dUnif_{\permutations_M}}$ and set $\smash{\AncAll \coloneqq \mathbf{B}^{\varSigma}}$. \label{alg:resampling_with_random_permutation:extra_step}
      \State Set $\smash{\StaRes^m \coloneqq \Sta^{A^m}}$ and $\smash{\weiRes^m \coloneqq W^{A^m} / \sum_{m=1}^M \resDistAlt^m(A^m)}$, for $m \in [M]$.
\end{algorithmic}
\end{algorithm}
\algorithmSepBelow
\end{figure}

Comparing Algorithm~\ref{alg:resampling_with_random_permutation} with Algorithm~\ref{alg:resampling}, we see that randomly permuting the the order of the ancestor indices only adds computational cost  Specifically, the former requires the additional Line~\ref{alg:resampling_with_random_permutation:extra_step}.

Sampling the index $K$ within the conditional version of $\resScheme$ becomes simpler compared to $\resSchemeAlt$ since the distribution of $K$ is now uniform under $\indDist(\ccdot|a)$. However, conditional resampling also requires sampling from $\smash{\resDist^k(\aux|\anc^k) \resDist^{-k}(\ancAll^{-k}|\anc^k, \aux) \delta_a(\anc^k)}$ and the random permutation makes the latter more involved:
\begin{align}
  \MoveEqLeft \resDist^k(\aux|\anc^k) \resDist^{-k}(\ancAll^{-k}|a^k, \aux) \delta_\anc(\anc^k)\\
  & = \frac{\resDist(\ancAll, \aux) \ind\{a^k = a\}}{\resDist^k(a)}\\
  & = \frac{\sum_{\mathbf{b} \in [N]^M}  \resDistAlt(\mathbf{b}, \aux) \sum_{\sigma \in \permutations_M} \dUnif_{\permutations_M}(\sigma) \delta_{\mathbf{b}^\sigma}(\ancAll)\ind\{a^k = a\}}{\frac{1}{M}\sum_{m=1}^M \resDistAlt^m(a)} \\
  & = \sum_{\mathbf{b} \in [N]^M} \sum_{l=1}^M  \frac{\resDistAlt^l(a)}{\sum_{m=1}^M \resDistAlt^m(a)} \resDistAlt^l(\aux|a)  \delta_{a}(b^l) \resDistAlt^{-l}(\mathbf{b}^{-l} |b^l, \aux) \smashoperator{\sum_{\sigma \in \permutations_M^k(l)}}  \dUnif_{\permutations_M^k(l)}(\sigma)
  \delta_{\mathbf{b}^\sigma}(\ancAll)\\
  & = \sum_{\mathbf{b} \in [N]^M}  \sum_{l=1}^M \kappa(l, \aux|a)  \delta_{a}(b^l)  \resDistAlt^{-l}(\mathbf{b}^{-l} | b^l, \aux) \smashoperator{\sum_{\sigma \in \permutations_M^k(l)}} \dUnif_{\permutations_M^k(l)}(\sigma)
   \delta_{\mathbf{b}^\sigma}(\ancAll),
\end{align}
where $\smash{\permutations_M^k(l) \coloneqq \{\sigma \in \permutations_M \mid \sigma(k) = l\}}
$ which, in particular, means that in the last two lines, $\smash{\sigma \in \permutations_M^k(l)}$ implies that $\smash{b^l} = \smash{b^{\sigma(k)} = a^{\sigma^{-1}(l)}} = \smash{a^k}$.

\begin{figure}[H]
\algorithmSepAbove
\begin{algorithm}[H]
\caption{conditional resampling with random permutation}\label{alg:conditional_resampling_with_random_permutation}
\begin{algorithmic}[1]
  \Require $a \in [N]$ with $\smash{W^a > 0}$.
    \State Sample $\smash{(L, \Aux, \mathbf{B}) \sim \kappa(l, \aux|a) \delta_{a}(b^l) \resDistAlt^{-l}(\mathbf{b}^{-l}|b^l, \aux)}$.
    \State Sample $\smash{K \sim \dUnif_{[M]}}$ as well as $\smash{\varSigma \sim \dUnif_{\permutations_M^K(L)}}$ and set $\smash{\AncAll \coloneqq \mathbf{B}^{\varSigma}}$. \label{alg:conditional_resampling_with_random_permutation:extra_step}
    \State Set $\smash{\StaRes^m \coloneqq \Sta^{A^m}}$ and $\smash{\weiRes^m \coloneqq W^{A^m} / \sum_{m=1}^M \resDistAlt^m(A^m)}$, for $m \in [M]$.
\end{algorithmic}
\end{algorithm}
\algorithmSepBelow
\end{figure}

Again by comparing Algorithm~\ref{alg:conditional_resampling_with_random_permutation} with Algorithm~\ref{alg:conditional_resampling}, we see that the random permutation incurs the additional Line~\ref{alg:conditional_resampling_with_random_permutation:extra_step}.

In summary, our work shows that randomly permuting the order of the ancestor indices is not needed (and only adds unnecessary computational cost) unless one wishes to exploit Proposition~\ref{prop:alternative_lambda_in_exchangeable_resampling_schemes} to implement conditional resampling in a way that always sets $K \coloneqq 1$ or $K \coloneqq M$, say, in Line~\ref{alg:conditional_resampling_with_random_permutation:extra_step}.

\subsubsection{Random cyclical shifts}

All the developments from the previous section remain valid if we replace the random permutations $\sigma \in \permutations_M$ by the subset of $\permutations_M$ which correspond to random cyclical shifts $\sigma \in \cycles_M \coloneqq \{\tau\colon [M] \to [M] \mid \exists\, s \in [M]\; \forall\, m \in [M]: \tau(m) = m + s \mod M\}$ (and if we similarly replace $\permutations_M^k(l)$ by $\cycles_M^k(l) \coloneqq \{\sigma \in \cycles_M \mid \sigma(k) = l\}
$). In particular, note that the resampling scheme $\resScheme$ induced by applying such random cyclical shifts to the order of the ancestor indices generated by $\resDistAlt$ still satisfies the assumption of Proposition~\ref{prop:alternative_lambda_in_exchangeable_resampling_schemes}.

\begin{figure}[H]
\algorithmSepAbove
\begin{algorithm}[H]
\caption{resampling with random shift}\label{alg:resampling_with_random_shift}
\begin{algorithmic}[1]
      \State Sample $\smash{(\mathbf{B}, \Aux) = (B^{1:M}, \Aux) \sim \resDistAlt}$.
      \State Sample $\smash{\varSigma \sim \dUnif_{\cycles_M}}$ and set $\smash{\AncAll \coloneqq \mathbf{B}^{\varSigma}}$. \label{alg:resampling_with_random_shift:extra_step}
      \State Set $\smash{\StaRes^m \coloneqq \Sta^{A^m}}$ and $\smash{\weiRes^m \coloneqq W^{A^m} / \sum_{m=1}^M \resDistAlt^m(A^m)}$, for $m \in [M]$.
\end{algorithmic}
\end{algorithm}
\algorithmSepBelow
\end{figure}

\begin{figure}[H]
\algorithmSepAboveAlt
\begin{algorithm}[H]
\caption{conditional resampling with random shift}\label{alg:conditional_resampling_with_random_shift}
\begin{algorithmic}[1]
  \Require $a \in [N]$ with $\smash{W^a > 0}$.
    \State Sample $\smash{(L, \Aux, \mathbf{B}) \sim \kappa(l, \aux|a) \delta_{a}(b^l) \resDistAlt^{-l}(\mathbf{b}^{-l}|b^l, \aux)}$. \label{alg:conditional_resampling_with_random_shift:1} 
    \State Sample $\smash{K \sim \dUnif_{[M]}}$ as well as $\smash{\varSigma \sim \dUnif_{\cycles_M^K(L)}}$ and set $\smash{\AncAll \coloneqq \mathbf{B}^{\varSigma}}$. \label{alg:conditional_resampling_with_random_shift:extra_step}
    \State Set $\smash{\StaRes^m \coloneqq \Sta^{A^m}}$ and $\smash{\weiRes^m \coloneqq W^{A^m} / \sum_{m=1}^M \resDistAlt^m(A^m)}$, for $m \in [M]$.
\end{algorithmic}
\end{algorithm}
\algorithmSepBelow
\end{figure}

Such random cyclical shifts were employed in \citet[][Algorithm~4]{chopin2015particle} to derive a conditional version of systematic resampling in which $K$ can always be set to $1$. Specifically,  \citet[][Algorithm~4]{chopin2015particle} can be derived from Algorithm~\ref{alg:conditional_resampling_with_random_shift} by taking $\resSchemeAlt$ to be the systematic resampling scheme and $a = 1$ as follows:
\begin{itemize}
  \item in Line~\ref{alg:conditional_resampling_with_random_shift:1}  of Algorithm~\ref{alg:conditional_resampling_with_random_shift}, sample
  \begin{align}
    \Aux \sim \frac{\lfloor MW^1\rfloor}{MW^1}\dUnif_{[0,1]}(u) + \biggl(1 - \frac{\lfloor MW^1\rfloor}{MW^1}\biggr) \dUnif_{(0, MW^1 - \lfloor MW^1\rfloor]}(u), \label{eq:conditional_systematic_resampling_with_random_shift_conditional_distribution_of_u_given_a}
  \end{align}
  which is justified by the fact that $\kappa(u|a)$ in
  \begin{align}
    \kappa(l, \aux|a)
    & = \kappa(u|a)\kappa(l|u,a)\\
    & = \biggl[\sum_{m=1}^M \frac{\vol(I^m(a))}{MW^a} \dUnif_{I^m(a)}(u)\biggr] \dUnif_{\{m \in [M] \mid u \in I^m(a)\}}(l),
  \end{align}
  simplifies to the r.h.s.\ in \eqref{eq:conditional_systematic_resampling_with_random_shift_conditional_distribution_of_u_given_a} due to $a = 1$;

  \item also in Line~\ref{alg:conditional_resampling_with_random_shift:1}  of Algorithm~\ref{alg:conditional_resampling_with_random_shift}, set $B^m \coloneqq F^{-1}([U + m - 1]/M)$, for $m \in [M]$, using Algorithm~\ref{alg:inverse_cdf_sampling} with $U^1 = \ldots = U^M = U$ and simply set $L \coloneqq 1$ both of which are justified by the fact that $a = 1$ and that given $u \sim \kappa(u|a)$,
  \begin{align}
     \delta_{a}(b^l) \resDistAlt^{-l}(\mathbf{b}^{-l}|b^l, \aux)
     & = \prod_{m=1}^M \delta_{F^{-1}([u + m - 1] / M)}(b^m)
  \end{align}
  does not depend on $l$;

  \item in Line~\ref{alg:conditional_resampling_with_random_shift:extra_step} of Algorithm~\ref{alg:conditional_resampling_with_random_shift},  appeal to Proposition~\ref{prop:alternative_lambda_in_exchangeable_resampling_schemes} to set $K \coloneqq 1$.
\end{itemize}

Furthermore, \citet[][Algorithms~5 \& 6]{karppinen2023conditional} introduced conditional versions of killing resampling and of systematic resampling with weight re-ordering based on such random cyclical shifts which could likewise be justified as instances of Algorithm~\ref{alg:conditional_resampling_with_random_shift} in which $\resSchemeAlt$ is killing resampling or systematic resampling with weight re-ordering, respectively.

Our work shows that such random cyclical shifts are only necessary if one wishes to exploit Proposition~\ref{prop:alternative_lambda_in_exchangeable_resampling_schemes} to implement conditional resampling in a way that always sets $K \coloneqq 1$ or $K \coloneqq M$, say, in Line~\ref{alg:conditional_resampling_with_random_shift:extra_step}.

\subsection{Relationship with $\alpha$SMC resampling/resamping matrices}
\label{app:subsec:alpha_smc_resampling}

The $\alpha$\gls{SMC} framework from \citet{whiteley2016role} resampling provides a matrix representation of resampling schemes for which $\spaceU \coloneqq \emptyset$ and the resampling distribution factorises as
\begin{align}
    \resDist(\ancAll) = \prod_{m=1}^M \resDist^m(\anc^m). \label{eq:alpha_smc_factorisation}
\end{align}
Let $\alpha = (\alpha^{m,n}) \in [0, \infty)^{M \times N}$ be some $\calF$-measurable \emph{resampling matrix} which is such that each row defines a probability measure on $[N]$, i.e., such that $\sum_{n=1}^N \alpha^{m,n} = 1$, for any $m \in [M]$. We will also assume that
\begin{align}
  \sum_{m=1}^M \alpha^{m,n} > 0, \label{eq:alpha_smc_support_condition}
\end{align}
for any $n \in [N]$ with $W^n > 0$.

Then $\alpha$\gls{SMC} represents $\rho^m(n)$ as
\begin{align}
  \resDist^m(n) \coloneqq \dCat\biggl(n; \biggl(\frac{\alpha^{m,k} W^k}{\sum_{l=1}^N \alpha^{m,l} W^l}\biggr)_{k = 1}^N \biggr),
\end{align}
which is well defined by \eqref{eq:alpha_smc_support_condition} (and the usual convention $\tfrac{0}{0} = 0$). 

Use of such resampling matrices goes back to at least \citet{hu2008basic}

To embed $\alpha$\gls{SMC} resampling within our framework, set
\begin{align}
    \lambda(m|n) \coloneqq \frac{\alpha^{m,n}}{\sum_{l=1}^M \alpha^{l,n}},
\end{align}
whenever $W^n > 0$, so that
\begin{align}
   \weiResFun(m,n) & = M W^n \frac{\lambda(m|n)}{\rho^m(n)}
   = M \sum_{l=1}^N \frac{\alpha^{m,l}}{\sum_{k=1}^M \alpha^{k,l}} W^l.
\end{align}
Then the $\alpha$\gls{SMC} resampling scheme $\resScheme$ is not necessarily balanced but still properly weighted and persistent.

The factorisation \eqref{eq:alpha_smc_factorisation} holds for most of the elementary resampling schemes from Section~\ref{subsec:elementary_resampling_schemes}, including trivial resampling, i.e., not resampling (recall that as explained in Remark~\ref{rem:integrating_out_auxiliary_variables_in_balanced_resampling_schemes}, whether or not we explicitly include the auxiliary variable $\Aux$ into the space is immaterial for balanced resampling schemes). And since the resampling matrix $\alpha$ is allowed to be some (measurable) function of the particle system, $\alpha$\gls{SMC} even certain kinds of \gls{ESS}-adaptive resampling.  Additionally, a conditional version of $\alpha$\gls{SMC} resampling was derived in \citet{huggins2019sequential}.

However, the $\alpha$\gls{SMC} framework does not cover resampling distributions which violate the factorisation from \eqref{eq:alpha_smc_factorisation}, such as the widely used (adaptive) systematic resampling scheme or more `exotic' schemes like chopthin resampling. This shows that our framework provides a strict generalisation of $\alpha$\gls{SMC}.

\section{Proofs for Section~\ref{sec:framework}}
\label{app:sec:proofs_for_section_2}

\begin{namedProof}[of Proposition~\ref{prop:alternative_lambda_in_exchangeable_resampling_schemes}]
  Since $\resScheme$ is balanced, $\weiResFun(m, n, \aux)$ is constant in $\aux$ so that
  \begin{align}
   f^b(\ancAll) 
   & \coloneqq \sum_{l=1}^M \frac{\weiResFun(l, \anc^l\!, \aux)}{\sum_{m=1}^M \weiResFun(m, \anc^m\!, \aux)} \ind\{a^l = b\}
   = \sum_{l=1}^M \frac{\Wei^{a^l} / \sum_{i=1}^M \resDist^i(a^l)}{\sum_{m=1}^M \Wei^{a^m} / \sum_{j=1}^M \resDist^j(a^m)} \ind\{a^l = b\},
  \end{align}
  is well defined and symmetric in all its $M$ arguments. Let $\mu(k, a, \ancAll, \diff \aux, b, l)$ be the extended distribution from the proof of Proposition~\ref{prop:proper_weighting_implies_invariance} and let $\mu'(k, a, \ancAll, \diff \aux, b, l)$ be the same distribution but with $\indDist(k,\diff \aux|a) = \indDist(k|a) \resDist^k(\diff \aux|a)$ replaced by $\indDist'(k|a)\resDist^k(\diff \aux|a)$. Then the marginal distribution $\mu(b)$ of $b$ under both these extended distributions is the same:
   \begin{align}
     \mu'(b)
     & = \sum_{k=1}^M W^a \indDist'(k|a) \E_{k,a}[f^b(\AncAll)] 
     = \E_{1,a}[f^b(\AncAll)]  = \sum_{k=1}^M W^a \indDist(k|a) \E_{k,a}[f^b(\AncAll)]
     = \mu(b).
 \end{align}
Thus, the proof of Proposition~\ref{prop:proper_weighting_implies_invariance} is not affected by the replacement of $\indDist(k|a)$ by $\indDist'(k|a)$. \hfill \qedwhite
\end{namedProof}

\section{Proofs for Section~\ref{sec:application_to_smc_methods}}

\subsection{CSMC with backward or ancestor sampling}
\label{app:subsec:csmc_with_backward_or_ancestor_sampling}

We again assume that $N_t \geq 2$ and make the following assumption on the mutation kernels to ensure that the transition densities $\smash{m_t(\ccdot|x) \coloneqq \diff M_t(\ccdot|x) / \diff \nu}$ 
are well defined for any $x \in \spaceX$ and $t \in [T]$.

\begin{assumption}\label{as:transition_density}
  There exists a measure $\nu$ on $\spaceX$ such that $M_t(\ccdot|x) \ll \nu$, for any $x \in \spaceX$, is absolutely continuous w.r.t.\ $nu$, for all $x \in \spaceX$ and $t \in [T]$.
\end{assumption}

Assumption~\ref{as:transition_density} guarantees the existence of the following \emph{backward kernel}, for any $k \in [N_t]$, any $t < T$ and with the notation $\smash{\staAll_t \coloneqq \sta^{1:N_t}}$:
\begin{align}
  \backwardKernel_t(k|\staAll_t, \sta_{t+1}; \WeiAll_t)
  & \coloneqq \dfrac{W_t^k G_{t+1}(\sta_t^k, \sta_{t+1}) m_{t+1}(\sta_{t+1}|\sta_t^k)}{\sum_{n=1}^{N_t} W_t^n G_{t+1}(\sta_t^n, \sta_{t+1}) m_{t+1}(\sta_{t+1}|\sta_t^n)}. \label{eq:backward_kernel}
\end{align}

\begin{figure}[htb]
\algorithmSepAboveAlt
\begin{algorithm}[H]
\caption{\gls{CSMC} with backward sampling}\label{alg:csmc_with_backward_sampling}
\begin{algorithmic}[1]
  \Require $\smash{\sta_{1:T} \in \spaceX^T}$ with $\smash{\pi_T(\sta_{1:T}) > 0}$.
  \For {$t = 1,\dotsc, T$}  \label{alg:csmc_with_backward_sampling:1a}
    \If{$t = 1$}
      \State Sample $\smash{K_1 \sim \indDist_1}$.
      \State Set $\smash{\weiRes_{t-1}^n \coloneqq \indDist_1(n)}$, for $n \in [N_1]$.
    \Else
      \State Sample $\smash{(K_t, \AncAll_{t-1}^{-K_t}, \Aux_{t-1}) \sim \indDist_t(k_t, \aux_{t-1}|K_{t-1};\WeiAll_{t-1}) \resDist_{t-1}^{-k_t}(\ancAll_{t-1}^{-k_t}| K_{t-1}, \aux_{t-1}; \WeiAll_{t-1})}$.
      \State Set $\smash{\Anc_{t-1}^{K_t} \coloneqq K_{t-1}}$.

           \State Set $\smash{\weiRes_{t-1}^n \coloneqq \weiResFun_t(n, \Anc_{t-1}^n, \Aux_{t-1}; \WeiAll_{t-1}) / N_t}$ via \eqref{eq:smc_post_resampling_weight_function}, and $\smash{\StaRes_{t-1}^n \coloneqq \Sta_{t-1}^{A_{t-1}^n}}$, for $n \in [N_t]$.
    \EndIf
      \State Set $\Sta_t^{K_t} \coloneqq \sta_t$, and sample $\Sta_t^n \sim M_t(\ccdot|\StaRes_{t-1}^n)$, for $n \in [N_t] \setminus \{K_t\}$.
      \State Set $\smash{w_t^n \coloneqq \weiRes_{t-1}^n G_t(\StaRes_{t-1}^n, \Sta_t^n)}$, and $\smash{W_t^n \coloneqq w_t^n / \sum_{m=1}^{N_t} w_t^m}$, for $n \in [N_t]$.  \label{alg:csmc_with_backward_sampling:1b}
   \EndFor
\State Sample $L_T \sim \dCat(\WeiAll_T)$ 
    and sample $L_t \sim \smash{\backwardKernel_t(\ccdot|\StaAll_t, \Sta_{t+1}^{L_{t+1}}; \WeiAll_t)}$,
    for $t = T-1,\dotsc, 1$.  \label{alg:csmc_with_backward_sampling:2}
    \State Return $\smash{\Sta_{1:T}' \coloneqq (\Sta_1^{L_1}, \dotsc, \Sta_T^{L_T})}$.  \label{alg:csmc_with_backward_sampling:4}
\end{algorithmic}
\end{algorithm}
\algorithmSepBelow
\end{figure}

\begin{figure}[H]
\algorithmSepAboveAlt
\begin{algorithm}[H]
\caption{\gls{CSMC} with ancestor sampling}\label{alg:csmc_with_ancestor_sampling}
\begin{algorithmic}[1]
  \Require $\smash{\sta_{1:T} \in \spaceX^T}$ with $\smash{\pi_T(\sta_{1:T}) > 0}$.
  \For {$t = 1,\dotsc, T$} \label{alg:csmc_with_ancestor_sampling:1a}
    \If{$t = 1$}
      \State Sample $\smash{K_1 \sim \indDist_1}$.
      \State Set $\smash{\weiRes_{t-1}^n \coloneqq \indDist_1(n)}$, for $n \in [N_1]$.
    \Else
      \State Sample $\smash{(K_t, \Aux_{t-1}) \sim \indDist_t(k_t, \aux_{t-1}|K_{t-1};\WeiAll_{t-1})}$.
      \State Sample $\smash{\AncAll_{t-1} \sim \backwardKernel_{t-1}(a_{t-1}^{K_t}|\StaAll_{t-1}, \sta_t; \WeiAll_{t-1})\resDist_{t-1}^{-K_t}(\ancAll_{t-1}^{-K_t}| \Anc_{t-1}^{K_t}, \Aux_{t-1}; \WeiAll_{t-1})}$,

           \State Set $\smash{\weiRes_{t-1}^n \coloneqq \weiResFun_t(n, \Anc_{t-1}^n, \Aux_{t-1}; \WeiAll_{t-1}) / N_t}$ via \eqref{eq:smc_post_resampling_weight_function}, and $\StaRes_{t-1}^n \coloneqq \Sta_{t-1}^{A_{t-1}^n}$, for $n \in [N_t]$.
    \EndIf
      \State Set $\Sta_t^{K_t} \coloneqq \sta_t$, and sample $\Sta_t^n \sim M_t(\ccdot|\StaRes_{t-1}^n)$ for $n \in [N_t] \setminus \{K_t\}$.
      \State Set $\smash{w_t^n \coloneqq \weiRes_{t-1}^n G_t(\StaRes_{t-1}^n, \Sta_t^n)}$, and $\smash{W_t^n \coloneqq w_t^n / \sum_{m=1}^{N_t} w_t^m}$, for $n \in [N_t]$. \label{alg:csmc_with_ancestor_sampling:1b}
   \EndFor
\State Sample $L_T \sim \dCat(\WeiAll_T)$ and set $\smash{L_t \coloneqq \Anc_t^{L_{t+1}}}$ for $t = T-1,\dotsc, 1$. \label{alg:csmc_with_ancestor_sampling:2}
    \State Return $\smash{\Sta_{1:T}' \coloneqq (\Sta_1^{L_1}, \dotsc, \Sta_T^{L_T})}$.  \label{alg:csmc_with_ancestor_sampling:4}
\end{algorithmic}
\end{algorithm}
\algorithmSepBelow
\end{figure}

The following result---proved in Appendix~\ref{app:subsec:proof_of_csmc_with_backward_or_ancestor_sampling_validity}---extends Proposition~\ref{prop:csmc_validity}.

\begin{proposition}\label{prop:csmc_with_backward_or_ancestor_sampling_validity}
  Assume that the resampling schemes $\resSchemeTime{t}$ are properly weighted and persistent for any $t \in [T] \setminus \{1\}$ and that $P$ is the Markov kernel induced by Algorithm~\ref{alg:csmc_with_backward_sampling} or Algorithm~\ref{alg:csmc_with_ancestor_sampling}. Then $P$ is $\smash{\pi_T}$-invariant.
\end{proposition}

\subsection{Extended state-space construction}

Throughout this section, we write $\smash{\StaAll_t \coloneqq \Sta_t^{1:N_t}}$ as well as $\mathbf{Z}_t \coloneqq (\StaAll_t, \AncAll_{t-1}, \Aux_{t-1})$, for $t > 1$ and $\mathbf{Z}_1 \coloneqq \StaAll_1$. The weights $\smash{\WeiAll_{t} = W_{t}^{1:N_{t}}}$ then depend on $\mathbf{Z}_{1:T}$ although we do not make this explicit to simplify the notation. The following distributions are well defined if the resampling schemes $\resSchemeTime{t}$ are all properly weighted and persistent.

\begin{enumerate}
\item The law of all the random variables generated by the \gls{SMC} algorithm (Algorithm~\ref{alg:smc}) is then
  \begin{align}
    \smc_T(\diff \mathbf{z}_{1:T})
    & \coloneqq \Biggl[ \prod_{n=1}^{N_1} M_1(\diff \sta_1^n)\biggr]  \prod_{t=2}^T \biggl[\resDist_{t-1}(\diff [\ancAll_{t-1}, \aux_{t-1}]; \WeiAll_{t-1}) \prod_{n=1}^{N_t} M_t(\diff \sta_t^n|\sta_{t-1}^{a_{t-1}^n})\biggr].
  \end{align}

\item Similarly, the law of all the random variables generated by the \gls{CSMC} algorithm with backward sampling (Algorithm~\ref{alg:csmc_with_backward_sampling})---under Assumption~\ref{as:transition_density}---is
\begin{align}
  \csmc_T(\diff[k_{1:T}, \mathbf{z}_{1:T}]|\sta_{1:T}) \backwardSampling_T(\diff[l_{1:T}, \sta_{1:T}']| \mathbf{z}_{1:T}),
\end{align}
where the first term is the law of all the random variables generated by Lines~\ref{alg:csmc_with_backward_sampling:1a}--\ref{alg:csmc_with_backward_sampling:1b}:
  \begin{align}
    \MoveEqLeft\csmc_T(\diff[k_{1:T}, \mathbf{z}_{1:T}]|\sta_{1:T})\\
    & \coloneqq \Biggl[ \indDist_1(\diff k_1) \delta_{\sta_1}(\diff \sta_1^{k_1}) \prodSubstackAligned{n}{=}{1}{n}{\neq}{k_1}^{N_1} M_1(\diff \sta_1^n)\Biggr] \prod_{t=2}^T \Biggl[ \indDist_t(\diff [k_t, \aux_{t-1}] | k_{t-1}, \WeiAll_{t-1})\\[-2ex]
    & \qquad \times \delta_{k_{t-1}}(\diff \anc_{t-1}^{k_t}) \resDist_{t-1}^{-k_t}(\diff \ancAll_{t-1}^{-k_t}  | \anc_{t-1}^{k_t}, \aux_{t-1}; \WeiAll_{t-1}) \delta_{\sta_t}(\diff \sta_t^{k_t}) \prodSubstackAligned{n}{=}{1}{n}{\neq}{k_t}^{N_t} M_t(\diff \sta_t^n|\sta_{t-1}^{a_{t-1}^n}) \Biggr];
  \end{align}
and the second term is the law of all the random variables generated in Lines~\ref{alg:csmc_with_backward_sampling:2}--\ref{alg:csmc_with_backward_sampling:4}:
\begin{align}
    \backwardSampling_T(\diff[l_{1:T}, \sta_{1:T}']| \mathbf{z}_{1:T})
    & \coloneqq W_T^{l_T} \biggl[ \prod_{t=1}^{T-1}
    \backwardKernel_t(\diff l_t|\staAll_t, \sta_{t+1}^{l_{t+1}}; \WeiAll_t)
    \biggr] \prod_{t=1}^T \delta_{\sta_t^{l_t}}(\diff \sta_t').
  \end{align}

  \item Finally, the law of all the random variables generated by the \gls{CSMC} algorithm with ancestor sampling (Algorithm~\ref{alg:csmc_with_ancestor_sampling})---again under Assumption~\ref{as:transition_density}--- is 
\begin{align}
  \csmcWithAncestorSampling_T(\diff[k_{1:T}, \mathbf{z}_{1:T}]|\sta_{1:T}) \ancestorTracing_T(\diff[l_{1:T}, \sta_{1:T}']| \mathbf{z}_{1:T}),
\end{align}
where the first term is the law of all the random variables generated by Lines~\ref{alg:csmc_with_ancestor_sampling:1a}--\ref{alg:csmc_with_ancestor_sampling:1b}:
  \begin{align}
    \MoveEqLeft\csmcWithAncestorSampling_T(\diff[k_{1:T}, \mathbf{z}_{1:T}]|\sta_{1:T})\\
    & \coloneqq \Biggl[ \indDist_1(\diff k_1) \delta_{\sta_1}(\diff \sta_1^{k_1}) \prodSubstackAligned{n}{=}{1}{n}{\neq}{k_1}^{N_1} M_1(\diff \sta_1^n)\Biggr]\\[-2ex]
    & \quad \times \prod_{t=2}^T \Biggl[ \indDist_t(\diff [k_t, \aux_{t-1}] | k_{t-1}, \WeiAll_{t-1})
    \backwardKernel_{t-1}(\diff a_{t-1}^{k_t}| \staAll_{t-1}, \sta_t; \WeiAll_{t-1})\\[-3ex]
    & \qquad\qquad \times \resDist_{t-1}^{-k_t}(\diff \ancAll_{t-1}^{-k_t}  | \anc_{t-1}^{k_t}, \aux_{t-1}; \WeiAll_{t-1}) \delta_{\sta_t}(\diff \sta_t^{k_t}) \prodSubstackAligned{n}{=}{1}{n}{\neq}{k_t}^{N_t} M_t(\diff \sta_t^n|\sta_{t-1}^{a_{t-1}^n}) \Biggr];
  \end{align}
and the second term is the law of all the random variables generated in Lines~\ref{alg:csmc_with_ancestor_sampling:2}--\ref{alg:csmc_with_ancestor_sampling:4}:
  \begin{align}
    \ancestorTracing_T(\diff[l_{1:T}, \sta_{1:T}']|\mathbf{z}_{1:T})
    \coloneqq W_T^{l_T} \biggl[ \prod_{t=1}^{T-1} \delta_{\anc_t^{l_{t+1}}}(\diff l_t) \biggr] \prod_{t=1}^T \delta_{\sta_t^{l_t}}(\diff \sta_t').
  \end{align}
\end{enumerate}

In the following, we use the convention that $\mu(\diff x) P(\diff y| x) = [\mu \otimes P](\diff [x, y])$ and with some abuse of notation regarding the order of the arguments. We also make the dependence of $\smash{\widehat{\normConst}_T = \widehat{\normConst}_T(\mathbf{Z}_{1:T})}$ on $\mathbf{Z}_{1:T}$ explicit in the notation.

\begin{lemma} \label{lem:radon--nikodym_derivative}
  Assume that the resampling schemes $\resSchemeTime{t}$ are properly weighted and persistent for all $t \in [T] \setminus \{1\}$. Then:
  \begin{enumerate}
    \item \label{lem:radon--nikodym_derivative:1} The Radon--Nikodym derivative
    \begin{align}
       \MoveEqLeft \dfrac{\diff [\gamma_T \otimes \csmc_T]}{\diff [\smc_T \otimes \ancestorTracing_T]}(\mathbf{z}_{1:T}, k_{1:T}, \sta_{1:T})\\
       & =  \widehat{\normConst}_T(\mathbf{z}_{1:T}) \ind\{(\sta_1^{k_1}, \dotsc, \sta_T^{k_T}) = \sta_{1:T}, (\anc_1^{k_2}, \dotsc, \anc_{T-1}^{k_T}) = k_{1:T-1}\},
    \end{align}
    is well defined for $[\smc_T \otimes \ancestorTracing_T]$-almost all $(\mathbf{z}_{1:T}, k_{1:T}, \sta_{1:T})$.

 \item \label{lem:radon--nikodym_derivative:2} The Radon--Nikodym derivative
    \begin{align}
       \dfrac{\diff [\gamma_T \otimes \csmcWithAncestorSampling_T]}{\diff [\smc_T \otimes \backwardSampling_T]}(\mathbf{z}_{1:T}, k_{1:T}, \sta_{1:T}) = \widehat{\normConst}_T(\mathbf{z}_{1:T}) \ind\{(\sta_1^{k_1}, \dotsc, \sta_T^{k_T}) = \sta_{1:T}\},
    \end{align}
    is well defined for $[\smc_T \otimes \backwardSampling_T]$-almost all $(\mathbf{z}_{1:T}, k_{1:T}, \sta_{1:T})$ if additionally Assumption~\ref{as:transition_density} holds.
  \end{enumerate}
\end{lemma}
\begin{namedProof}
  For Part~\ref{lem:radon--nikodym_derivative:1}, since the resampling schemes $\resSchemeTime{t}$ are properly weighted and persistent, for $[\smc_T \otimes \ancestorTracing_T]$-almost all $(\mathbf{z}_{1:T}, k_{1:T}, \sta_{1:T})$, we have
\begin{align}
  \MoveEqLeft \dfrac{\diff [\gamma_T \otimes \csmc_T]}{\diff [\smc_T \otimes \ancestorTracing_T]}(\mathbf{z}_{1:T}, k_{1:T}, \sta_{1:T})\\
  & = \indDist_1(k_1) \biggl[\prod_{t=1}^T G_t(\sta_{t-1}^{k_{t-1}}, \sta_t^{k_t})\biggr] (W_T^{k_T})^{-1}\\[-0.5ex]
  & \quad \times \biggl[ \prod_{t=2}^{T} \frac{\indDist_t(k_t, \aux_{t-1} | k_{t-1}, \WeiAll_{t-1}) \resDist_{t-1}^{-k_t}(\ancAll_{t-1}^{-k_t}  | \anc_{t-1}^{k_t}, \aux_{t-1}; \WeiAll_{t-1})}{\resDist_{t-1}(\ancAll_{t-1}, \aux_{t-1} ; \WeiAll_{t-1})}\biggr]\\
  & \quad \times \ind\{(\sta_1^{k_1}, \dotsc, \sta_T^{k_T}) = \sta_{1:T}, (\anc_1^{k_2}, \dotsc, \anc_{T-1}^{k_T}) = k_{1:T-1}\}\\
  & = \indDist_1(k_1) G_1(\sta_1^{k_1}) (W_T^{k_T})^{-1} \biggl[ \prod_{t=2}^{T} G_t(\sta_{t-1}^{k_{t-1}}, \sta_t^{k_t}) \frac{\indDist_t(k_t, \aux_{t-1} | k_{t-1}, \WeiAll_{t-1}) }{\resDist_{t-1}^{k_t}(\anc_{t-1}^{k_t}, \aux_{t-1} ; \WeiAll_{t-1})}\biggr]\\[-0.5ex]
  & \quad \times \ind\{(\sta_1^{k_1}, \dotsc, \sta_T^{k_T}) = \sta_{1:T}, (\anc_1^{k_2}, \dotsc, \anc_{T-1}^{k_T}) = k_{1:T-1}\}\\
  & = \widehat{\normConst}_T(\mathbf{z}_{1:T}) \ind\{(\sta_1^{k_1}, \dotsc, \sta_T^{k_T}) = \sta_{1:T}, (\anc_1^{k_2}, \dotsc, \anc_{T-1}^{k_T}) = k_{1:T-1}\},
\end{align}
where the last identity uses that $\smash{\indDist_1(k_1) G_1(\sta_1^{k_1}) = w_1^{k_1}}$ and that the terms within the brackets in the penultimate line are equal to $\smash{w_t^{k_t} / W_{t-1}^{k_{t-1}}}$.

 Part~\ref{lem:radon--nikodym_derivative:2} follows analogously using the definition of the backward kernels in \eqref{eq:backward_kernel}. \hfill \qedwhite
\end{namedProof}

\subsection{Proof of Proposition~\ref{prop:smc_validity}}
\label{app:subsec:proof_of_smc_validity}

The proof slightly extends the `\gls{SMC} as importance sampling' argument from \citet[][Appendix~B1]{andrieu2010particle} to our setting.

By Part~\ref{lem:radon--nikodym_derivative:1} of Lemma~\ref{lem:radon--nikodym_derivative}, we have $\gamma_T \otimes \csmc_T \ll \smc_T \otimes \ancestorTracing_T$. The proof then follows by a change of measure and the tower property of conditional expectation:
\begin{align}
  \gamma_T(\testFun)
  & = \int \testFun(\sta_{1:T}) [\gamma_T \otimes \csmc_T](\diff[\mathbf{z}_{1:T}, k_{1:T}, \sta_{1:T}])\\
  & = \E\biggl[\testFun(\Sta_{1:T}) \dfrac{\diff [\gamma_T \otimes \csmc_T]}{\diff [\smc_T \otimes \ancestorTracing_T]}(\mathbf{Z}_{1:T}, K_{1:T}, \Sta_{1:T})\biggr]\\
  & =  \E\biggl[\E\biggl(\testFun(\Sta_{1:T}) \dfrac{\diff [\gamma_T \otimes \csmc_T]}{\diff [\smc_T \otimes \ancestorTracing_T]}(\mathbf{Z}_{1:T}, K_{1:T}, \Sta_{1:T})\bigg|\mathbf{Z}_{1:T}\biggr)\biggr]\\
  & = \E\biggl[\widehat{\normConst}_T(\mathbf{Z}_{1:T}) \sum_{n=1}^{N_T} W_T^n \testFun(\Sta_{1:T}^{(n)}) \biggr]\\
  & =  \E[ \hat{\gamma}_T(\testFun)],
\end{align}
where the expectations in the first three lines are w.r.t.\ $(\mathbf{Z}_{1:T}, K_{1:T}, \Sta_{1:T}) \sim \smc_T \otimes \ancestorTracing_T$ and the expectations in the last two lines are w.r.t.\ $\smash{\mathbf{Z}_{1:T} \sim \smc_T}$. \hfill \qedwhite

\subsection{Proof of Proposition~\ref{prop:csmc_validity}}
\label{app:subsec:proof_of_csmc_validity}

The first part of Lemma~\ref{lem:radon--nikodym_derivative}, gives
\begin{align}
    \MoveEqLeft \pi_T(\diff \sta_{1:T}) \csmc_T(\diff[k_{1:T}, \mathbf{z}_{1:T}]|\sta_{1:T}) \ancestorTracing_T(\diff[l_{1:T}, \sta_{1:T}']| \mathbf{z}_{1:T})\\
    & = \widehat{\normConst}_T(\mathbf{z}_{1:T}) \smc_T(\diff \mathbf{z}_{1:T}) \ancestorTracing_T(\diff[k_{1:T}, \sta_{1:T}]|\mathbf{z}_{1:T}) \ancestorTracing_T(\diff[l_{1:T}, \sta_{1:T}']|\mathbf{z}_{1:T})\\
    & = \pi_T(\diff \sta_{1:T}') \csmc_T(\diff[l_{1:T}, \mathbf{z}_{1:T}]|\sta_{1:T}') \ancestorTracing_T(\diff[k_{1:T}, \sta_{1:T}]| \mathbf{z}_{1:T}).
\end{align}
This completes the proof. \hfill \qedwhite

\subsection{Proof of Proposition~\ref{prop:csmc_with_backward_or_ancestor_sampling_validity}}
\label{app:subsec:proof_of_csmc_with_backward_or_ancestor_sampling_validity}

The proof could be established in a number of ways, e.g., using partially-collapsed Gibbs sampling arguments \citep{whiteley2010particle, lindsten2014particle}. Here, we use the proof technique from \citet{finke2015extended} which is convenient in the sense that it simultaneously guarantees the validity of backward and ancestor sampling.

Specifically, appealing to both parts of Lemma~\ref{lem:radon--nikodym_derivative},
\begin{align}
    \MoveEqLeft \pi_T(\diff \sta_{1:T}) \csmc_T(\diff[k_{1:T}, \mathbf{z}_{1:T}]|\sta_{1:T}) \backwardSampling_T(\diff[l_{1:T}, \sta_{1:T}']| \mathbf{z}_{1:T})\\
    & = \widehat{\normConst}_T(\mathbf{z}_{1:T}) \smc_T(\diff \mathbf{z}_{1:T}) \ancestorTracing_T(\diff[k_{1:T}, \sta_{1:T}]|\mathbf{z}_{1:T}) \backwardSampling_T(\diff[l_{1:T}, \sta_{1:T}']|\mathbf{z}_{1:T})\\
    & = \pi_T(\diff \sta_{1:T}') \csmcWithAncestorSampling_T(\diff[l_{1:T}, \mathbf{z}_{1:T}]|\sta_{1:T}') \ancestorTracing_T(\diff[k_{1:T}, \sta_{1:T}]| \mathbf{z}_{1:T}).
\end{align}
This completes the proof in the case of backward or ancestor sampling. \hfill \qedwhite

\end{document}